\documentclass[acmtog]{acmart}
\acmSubmissionID{159}

\usepackage{booktabs} %

\usepackage{booktabs}
\usepackage{multirow}
\usepackage[table]{xcolor} %

\usepackage[dvipsnames]{xcolor}

\newcommand{\method}{{{ARDY}}\xspace}

\usepackage[ruled]{algorithm2e} %

\SetAlFnt{\small}
\SetAlCapFnt{\small}
\SetAlCapNameFnt{\small}
\SetAlCapHSkip{0pt}

\AtEndPreamble{
    \usepackage[capitalize]{cleveref}
    \crefname{section}{Sec.}{Secs.}
    \Crefname{section}{Section}{Sections}
    \Crefname{table}{Table}{Tables}
    \crefname{table}{Tab.}{Tabs.}
    \crefname{appendix}{App.}{Apps.}
    \Crefname{appendix}{Appendix}{Appendices}
}

\makeatletter
\DeclareRobustCommand\onedot{\futurelet\@let@token\@onedot}
\def\@onedot{\ifx\@let@token.\else.\null\fi\xspace}

\def\eg{\emph{e.g}\onedot} 
\def\ie{\emph{i.e}\onedot}

\makeatother

\usepackage{booktabs}
\usepackage{pifont}  %
\usepackage{xcolor}

\newcommand{\cmark}{\textcolor{green!60!black}{\ding{51}}} %
\newcommand{\xmark}{\textcolor{red}{\ding{55}}}           %

\acmJournal{TOG}
\acmVolume{45}
\acmNumber{4}
\acmArticle{86}
\acmYear{2026}
\acmMonth{7}

\setcopyright{cc}
\setcctype{by}
\acmJournal{TOG}
\acmYear{2026} \acmVolume{45} \acmNumber{4} \acmArticle{86}
\acmMonth{7} \acmDOI{10.1145/3811284}

\begin{document}
\title{ARDY: Autoregressive Diffusion with Hybrid Representation for Interactive Human Motion Generation}

\author{Kaifeng Zhao}
\orcid{0000-0002-7278-3329}
\email{kaifeng.zhao@inf.ethz.ch}
\affiliation{%
  \institution{NVIDIA}
  \country{Switzerland}
}
\affiliation{%
  \institution{ETH Zürich}
  \country{Switzerland}
}

\author{Mathis Petrovich}
\orcid{0000-0002-0859-1170}
\email{mpetrovich@nvidia.com}
\affiliation{%
  \institution{NVIDIA}
  \country{Switzerland}
}

\author{Haotian Zhang}
\orcid{0009-0008-0293-337X}
\email{haotianz@nvidia.com}
\affiliation{%
  \institution{NVIDIA}
  \country{USA}
}

\author{Tingwu Wang}
\orcid{0000-0003-2006-0660}
\email{tingwuw@nvidia.com}
\affiliation{%
  \institution{NVIDIA}
  \country{USA}
}

\author{Siyu Tang}
\orcid{0000-0002-1015-4770}
\email{siyu.tang@inf.ethz.ch}
\affiliation{%
  \institution{ETH Zürich}
  \country{Switzerland}
}

\author{Davis Rempe}
\orcid{0000-0003-2256-5507}
\email{drempe@nvidia.com}
\affiliation{%
  \institution{NVIDIA}
  \country{USA}
}

\renewcommand\shortauthors{Zhao et al.}

\begin{abstract}
Generating realistic 3D human motions in real-time within interactive applications is key for animation, simulation, and humanoid robotics.
While recent offline motion generation approaches offer precise control via text and kinematic constraints, they lack the inference speed required for interactive settings. Conversely, existing online methods enable real-time synthesis but often sacrifice controllability or struggle with complex text semantics and long-horizon goals due to limited context windows. 
In this work, we introduce ARDY, a streaming generation framework that bridges this gap by enabling high-fidelity motion generation controllable via online text prompts and flexible kinematic constraints.
ARDY employs a hybrid representation that combines explicit root features with a latent body embedding, balancing precise trajectory control with efficient generative learning.
We propose a two-stage autoregressive transformer denoiser that features variable history context and supports conditioning on flexible, long-horizon kinematic constraints.
By training on a large-scale motion capture dataset and being directly conditioned on text labels and kinematic constraints sampled from ground truth poses, ARDY natively learns controllable generation that supports online prompting and flexible long-horizon goals.
Extensive evaluations on the HumanML3D benchmark and the large-scale, high-fidelity Bones Rigplay dataset demonstrate ARDY's high motion quality and constraint adherence, validating the efficacy of our key architectural decisions.
Finally, we demonstrate the method’s practical versatility through an interactive demo featuring dynamic text control, diverse keyframe pose constraints, path following, and interactive locomotion control via mouse and keyboard.
Supplementary video results, code, and model releases can be found at \url{https://research.nvidia.com/labs/sil/projects/ardy/}.
\end{abstract}



\begin{teaserfigure}
    \centering
    \includegraphics[trim={0pt 0pt 0pt 0pt}, clip,width=\linewidth]{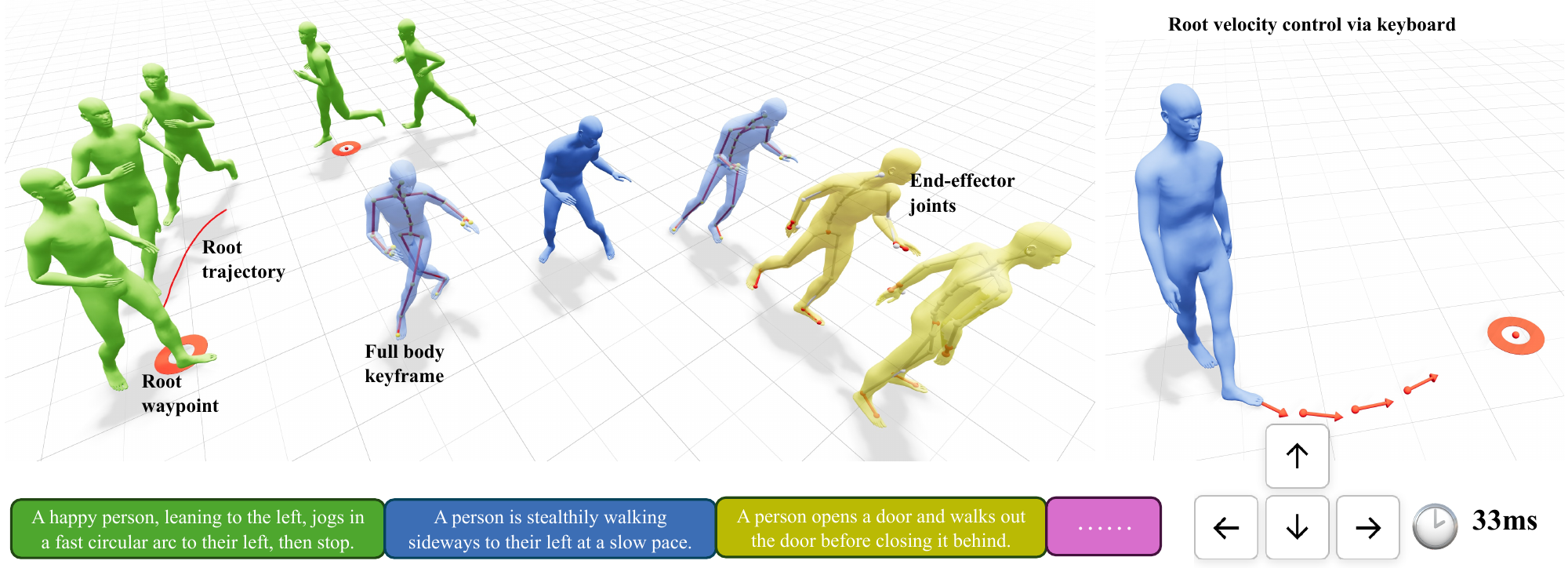}
    \caption{We present \textbf{\method}, an autoregressive diffusion model designed for interactive human motion generation. Our approach natively supports online text prompting alongside a comprehensive suite of flexible kinematic constraints — including root waypoints and trajectories, full-body keyframes, and sparse joint positions and rotations — over long horizons. \method enables controllable and responsive interactive motion synthesis from real-time user inputs such as mouse and keyboard commands, with our efficient 4-step diffusion model achieving an average generation latency of 33~ms.
    }
\label{fig:teaser}
\end{teaserfigure}

\maketitle

\section{Introduction}
\label{sec:intro}

Learning to generate realistic 3D human motions has become a promising direction with applications ranging from character animation and simulation to humanoid robotics. 
Offline authoring models can benefit animators and game developers through intuitive controls like text and kinematic constraints~\cite{xie2024omnicontrol,pinyoanuntapong2025MaskControl}.
Meanwhile, interactive motion generators~\cite{shi2024amdm,xiao2025motionstreamer} are key for characters in games and simulations to react to their environment and user inputs in real time. 
Besides digital humans, recent work in real-world humanoid robot control~\cite{liao2025beyondmimic,he2025asap,zhao2025resmimic, luo2025sonic} relies heavily on high-quality human motions for supervision during training or planning at runtime.

Recent methods in \emph{offline} motion modeling generate a full sequence of poses in parallel. Modern generative models such as diffusion~\cite{tevet2023mdm,zhang2024motiondiffuse, karunratanakul2023gmd, Kimodo2026} and generative masked modeling~\cite{jiang2024motiongpt,guo2024momask, pinyoanuntapong2025MaskControl} allow synthesized motions to follow complex text prompts and kinematic constraints such as pose keyframes and joint positions. 
While these methods are expressive and controllable, their spatiotemporal design and/or slow inference time are usually not suitable for interactive applications such as computer games or robot control.

In contrast, \emph{online} models generate motion at runtime~\cite{holden2017phase,ling2020character,chen2024camdm}, usually in an autoregressive fashion. 
While these models are fast and capable of producing realistic animations, they tend to sacrifice controllability. Some approaches support text conditioning but lack kinematic control~\cite{xiao2025motionstreamer}, while others enable kinematic constraints but can not accept text input~\cite{shi2024amdm,chen2024camdm}.
Although a few recent methods integrate both text and kinematic constraints control~\cite{zhao2025DartControl,tevet2025closd}, their restricted context windows limit the understanding of global text semantics and the execution of long-horizon kinematic goals.

In this work, we aim to get the best of both: controllability through complex text prompts and flexible kinematic goal constraints, while generating motion in a streaming fashion that enables online interactivity (see \cref{fig:teaser}).
To achieve this, we introduce \textbf{\method}, an \textbf{A}uto-\textbf{R}egressive \textbf{D}iffusion model that leverages a h\textbf{Y}brid pose representation to generate high-quality motion interactively, conditioned on online text prompts and flexible kinematic constraints from user inputs.
\method is comprised of two main components. 
First, \method employs a hybrid motion representation that decomposes motion into an explicit root feature and a latent body embedding derived from a learned tokenizer. This hybrid representation enables explicit and accurate root control during generation while maintaining a compact representation for efficient generative learning.
Second, \method utilizes an autoregressive transformer denoiser for interactive motion generation, conditioned on a text prompt and kinematic constraints that can be spatiotemporally sparse and span long horizons.
To handle variable and potentially sparse constraints, we represent the constraints as a masked motion sequence that is injected as input conditioning to the autoregressive denoiser.
The denoiser features a variable history context and supports kinematic goals extending beyond a single generation window, which are essential for complex long-term motion semantics and long-horizon kinematic goal reaching.
Moreover, the autoregressive denoiser employs an interleaved two-stage architecture: it first predicts the clean explicit root, then predicts the clean latent body embedding conditioned on the first-stage root prediction. These two stages operate in an interleaved manner within the denoising loop, ensuring continuous mutual influence between root and body motion. This staged design is crucial for simultaneously satisfying text instructions and kinematic constraints.
By training on a large-scale dataset with text labels and kinematic constraints sampled from the ground truth motion itself, \method learns conditional generation that supports online prompting and long-horizon kinematic goals, eliminating the need for additional control modules~\cite{shi2024amdm,zhao2025DartControl,pinyoanuntapong2025MaskControl} such as expensive test-time optimization or RL-based control policies.

We present an interactive demo that highlights the practical capabilities of our method, including dynamic text control, dense and sparse key-pose constraints, path following, and real-time locomotion control via mouse and keyboard. This demonstration showcases the potential for generative models to power next-generation interactive animation systems.
Moreover, we validate our design choices on the Bones Rigplay~\cite{BonesStudio2026} dataset—featuring a significantly larger scale and higher quality than the public HumanML3D dataset—to assess the impact of key architectural decisions.
Furthermore, we evaluate \method against state-of-the-art offline and autoregressive conditional motion generation methods on the public HumanML3D~\cite{guo2022humanml3d} benchmark, validating its strong motion quality and kinematic constraint adherence in a controlled setting that isolates the effects of proprietary data. 

In summary, the key contributions of this paper are (1) a hybrid latent-body explicit-root representation amenable to fast and controllable motion generation, (2) a two-stage autoregressive diffusion model featuring variable history context length and support for long-horizon kinematic constraint conditioning, including full-body keyframes, root waypoints, root paths, and end-effector positions/rotations, and (3) an extensive evaluation on a large-scale, production-quality dataset that highlights the efficacy of our design choices and demonstrates the strong capabilities of \method.

\section{Related Work}
\label{sec:relwork}

\begin{table*}[t]
\centering
\caption{\textbf{Method Feature Comparison.} Comparison of the proposed \method with existing conditional 3D motion generation methods. We delineate various capabilities including real-time performance, online prompting, supported spatial control types, the architectural mechanism of control (\ie, whether each method requires test-time optimization or RL policies), and the maximum history and future context length in model generation.}
\label{table:comparison}
\scriptsize

\begin{tabular*}{\textwidth}{l@{\extracolsep{\fill}}ccccccccc}
\toprule

\multirow{2}{*}{\textbf{Method}} & 
\multirow{2}{*}{\shortstack{\textbf{Real-time} \\ \textbf{generation}}} & 
\multirow{2}{*}{\shortstack{\textbf{Online text} \\ \textbf{prompting}}} &
\multicolumn{3}{c}{\textbf{Spatial control}} & 
\multicolumn{2}{c}{\textbf{Native control}} & 
\multicolumn{2}{c}{\textbf{Context length (s)}} \\ 

\cmidrule(lr){4-6} \cmidrule(lr){7-8} \cmidrule(lr){9-10}

 & & & 
 \shortstack{Root trajectory} & 
 \shortstack{Joint position} & 
 \shortstack{Joint rotation} & 
 \shortstack{No optimization} & 
 \shortstack{No RL policy} &
 History & 
 Future \\
\midrule

MaskControl \cite{pinyoanuntapong2025MaskControl}   & \xmark & \xmark & \cmark & \cmark & \xmark & \xmark & \cmark & N/A & 10.00 \\
Kimodo \cite{Kimodo2026}   & \xmark & \xmark & \cmark & \cmark & \cmark & \cmark & \cmark & N/A & 10.00 \\
AMDM \cite{shi2024amdm}          & \cmark & \xmark & \cmark & \cmark & \xmark & \cmark & \xmark & 0.03 & 0.03 \\
CAMDM  \cite{chen2024camdm}        & \cmark & \xmark & \cmark & \xmark & \xmark & \cmark & \cmark & 0.33 & 1.50 \\
MotionStreamer \cite{xiao2025motionstreamer}& \cmark & \cmark & \xmark & \xmark & \xmark & N/A & N/A & 10.00 & 10.00 \\
DartControl  \cite{zhao2025DartControl}  & \cmark & \cmark & \cmark & \cmark & \xmark & \xmark & \xmark & 0.07 & 0.27 \\
DiP   \cite{tevet2025closd}           & \cmark & \cmark & \cmark & \cmark & \xmark & \cmark & \cmark & 1.00 & 2.00 \\
\textbf{\method (Ours)}  & \cmark & \cmark & \cmark & \cmark & \cmark & \cmark & \cmark & 8.00 & 10.00 \\
\bottomrule
\end{tabular*}
\end{table*}

In this section, we summarize relevant work in conditional 3D human motion generation and how our method fits in context. 
For this purpose, we define \textit{offline} motion generation as a method that generates a full spatiotemporal sequence of poses in parallel, while \textit{online/interactive/runtime/streaming} motion generation refers to an autoregressive method that generates poses sequentially (either individually or in chunks) and can therefore react to dynamically changing conditions (\eg, new text prompts or constraints).

\paragraph{Offline Human Motion Generation}
A primary focus of many recent offline motion generation works is text conditioning.
Enabled by motion datasets with natural language descriptions~\cite{plappert2016kit}, early work on this problem employed VAE-based architectures for diverse generation~\cite{petrovich22temos,guo2022humanml3d}.
More recently, diffusion models have proven to be effective at capturing the complex distribution of text and motion, enabling high-quality motion generation from prompts~\cite{tevet2023mdm,chen2023executing,zhang2024motiondiffuse}. 
Motion diffusion models are also capable of flexible kinematic control, enabling ``any-joint-any-time'' constraints on generated motions~\cite{xie2024omnicontrol,karunratanakul2024optimizing,karunratanakul2023gmd, Kimodo2026}. However, the iterative denoising process for potentially long motions tends to be too slow for interactive applications. Some methods have considerably sped up the denoising process by reducing the number of required steps~\cite{dai2025motionlcm,zhou2024emdm}, but are still designed to generate all poses in parallel. %
While some diffusion approaches can handle a temporal sequence of input prompts, these methods generate all prompts jointly offline~\cite{barquero2024flowmdm,petrovich24stmc,li2025unimotion}, which is not suitable for interactive applications. 

Another line of work leverages a discrete tokenized representation of human motion.
Methods like MoMask~\cite{guo2024momask} and MMM~\cite{pinyoanuntapong2024mmm} generate motion from text by training a VQ-VAE motion tokenizer followed by a masked transformer that iteratively predicts masked poses, eventually resulting in a latent motion that can be decoded~\cite{meng2025rethinking,pinyoanuntapong2024bamm}.
Some tokenized approaches also support precise kinematic controls through test-time-optimization~\cite{wan2024tlcontrol,pinyoanuntapong2025MaskControl}.
Besides masked models, several approaches take inspiration from language models~\cite{radford2018gpt} and use autoregressive transformers to generate a sequence of motion tokens that are decoded to human poses~\cite{zhang2023generating,jiang2024motiongpt,fan2025gotozero,lu2025scamo}. 
While these methods are in fact autoregressive, they are generally large and slow models, designed for offline motion generation without support for precise kinematic control.

Our method \method delivers text-following and kinematic control capabilities on par with recent offline models, while operating within an interactive framework. This is achieved through a novel two-stage diffusion architecture that denoises a hybrid combination of latent (tokenized) body and explicit root representations.

\paragraph{Interactive Motion Generation}
Early works in autoregressive motion modeling leveraged non-linear latent variable models~\cite{taylor2006modeling} and recurrent neural networks~\cite{fragkiadaki2015recurrent}.
Non-generative autoregressive prediction models~\cite{holden2017phase,starke2019neural,starke2022deepphase} have been trained for reactive character control by conditioning on various combinations of past and future poses and trajectory information. 
In parallel, data-driven interactive animation systems such as Learned Motion Matching~\cite{holden2020learned} and Control Operators~\cite{gou2025control} enable responsive real-time character control via learned similarity metrics and modular control primitives rather than explicit generative modeling.
Moving into generative approaches, autoregressive VAE models learned a low-dimensional motion latent space for task-based RL control~\cite{ling2020character,zhang2022wanderings} and tracking via optimization~\cite{rempe2021humor}.
Similar approaches have learned human-object interactions~\cite{starke2019neural,hassan2021stochastic,zhao2023synthesizing} by conditioning the model on object geometry in addition to the future pose information.

Autoregressive motion diffusion models have taken the approaches developed for offline generation and made them amenable to interactive settings, primarily through shorter motion generation horizon and fewer denoising steps~\cite{shi2024amdm,chen2024camdm,zhang2024tedi,ji2025towards,zhang2025primal,jiang2024autonomous,zhao2025DartControl,wu2025uniphys}.
A-MDM learns to denoise the next pose in a motion given the previous pose, and allows flexible kinematic constraints through inpainting or RL control~\cite{shi2024amdm}. 
Similarly, CAMDM~\cite{chen2024camdm} and PRIMAL~\cite{zhang2025primal} denoise a small window of future frames given a handful of past frames. CAMDM is conditioned on a future trajectory to follow while PRIMAL relies on guidance and an additional ControlNet for velocity, heading, and waypoint control. 
While CAMDM and PRIMAL show action label conditioning, none of these methods support complex text prompting.
UniPhys~\cite{wu2025uniphys} enables text control, but relies entirely on test-time guidance for kinematic controls, which is inefficient for interactive applications.
Closest to our work is DiP~\cite{tevet2025closd}, which extends CAMDM by adding conditioning on text and 3D target joint locations provided every two seconds. However, DiP's short history and prediction horizon limit its ability to handle complex text prompts that require longer history context, and prevent it from satisfying kinematic constraints beyond its short generation horizon.

Latent diffusion has also been leveraged for interactive motion generation~\cite{zhao2025DartControl,xiao2025motionstreamer,cen2025ready_to_react}.
DartControl~\cite{zhao2025DartControl} uses a VAE to learn a continuous latent representation of motion primitives, then a diffusion model that predicts future motion in this latent space. Similar to DiP, DartControl is limited by a short history context, and kinematic control such as 2D waypoint reaching or full-body in-betweening requires test-time-optimization or training an additional RL control policy.
MotionStreamer~\cite{xiao2025motionstreamer} also learns a continuous latent space using a causal convolutional autoencoder, then trains a causal transformer denoiser to generate the next latent conditioned on the past and text input. Similar to our approach, MotionStreamer is trained on variable history length, making it more robust to complex prompts. However, it lacks support for kinematic goal constraints.

Several autoregressive diffusion models have been paired with physics-based controllers to carry out generated motions in simulation~\cite{tevet2025closd,wu2025uniphys,huang2025diffusecloc,rempe2023trace,ren2023insactor}. 
Fully physics-based runtime character control is also an active area of study~\cite{peng2022ase,luo2023universal}, which has recently enabled both kinematic control and preliminary text prompting~\cite{tessler2024maskedmimic,wu2025uniphys}. 

As shown in \cref{table:comparison}, our approach enables real-time generation with native support for online text prompting, variable-length history contexts, and flexible long-horizon kinematic constraints—a combination of capabilities unmatched by prior works.

\section{Method: \method}
\label{sec:method}
Our method \method consists of two main components: (1) a motion tokenizer first learns a compact latent representation of body motion, and then (2) an autoregressive two-stage motion diffusion model learns to denoise hybrid motion tokens containing latent body motion and explicit root motion.
Our hybrid representation is introduced in \cref{sec:hybrid_rep} followed by the body motion tokenizer in \cref{sec:tokenizer}. The autoregressive generation problem formulation is detailed in \cref{sec:problem_formulation} and then the diffusion model that solves it is described in \cref{sec:diffusion}.
Finally, \cref{sec:impl_details} covers implementation details.

\subsection{Hybrid Motion Representation}
\label{sec:hybrid_rep}
To balance the representational compactness required for efficient generative learning with the need for direct, precise control via explicit feature overwriting,
we propose a hybrid motion representation that decouples root motion from body motion. Specifically, root trajectories are represented in an explicit, interpretable form, while body motion is encoded in a compact latent space.
In this section, we give a high-level overview of the hybrid motion representation and its advantages for generation before detailing how the latent component is learned in \cref{sec:tokenizer}.

\paragraph{Explicit Motion Representation}
Our hybrid representation builds on an explicit motion representation, which we describe first for context. 
Each frame of a motion that uses this explicit representation $\mathbf{m} = (\mathbf{m}_\text{root}, \mathbf{m}_\text{body}) \in \mathbb{R}^M$ is defined as a tuple of root and body skeleton joint features 
\begin{equation}
    \mathbf{m}_\text{root} = (\mathbf{p}, \cos{\psi}, \sin{\psi}) \in \mathbb{R}^5, \quad \mathbf{m}_\text{body} = (\boldsymbol{\theta}, \mathbf{J}, \dot{\mathbf{J}}, \mathbf{c}),
\label{eqn:explicit_rep}
\end{equation}
where $\mathbf{p} \in \mathbb{R}^3$ denotes the global root position, $\psi \in (-\pi, \pi]$ denotes the root heading angle, $\boldsymbol{\theta} \in \mathbb{R}^{6j}$ denotes the 6D representation \cite{zhou2019continuity} of the global joint rotations for all $j$ skeleton joints including the root, $\mathbf{J} \in \mathbb{R}^{3j-3}$ denotes the non-root joint positions subtracted by the planar root position, $\dot{\mathbf{J}} \in \mathbb{R}^{3j}$ denotes the global joint velocities, and $\mathbf{c}  \in \mathbb{R}^{4}$ denotes the binary floor contact label for the feet joints.
The explicit representation feature size $M$ depends on the number of joints in the skeleton.

\paragraph{Hybrid Motion Representation}
Our hybrid motion representation is formed by simply replacing the body component of the pose feature with a latent embedding.
Concretely, a single pose $\mathbf{x}$ of a motion using the hybrid representation is a tuple
\begin{equation}
    \mathbf{x} = ( \mathbf{m}_\text{root}, \mathbf{x}_\text{body}  )
\end{equation}
where $\mathbf{x}_\text{body} \in \mathbb{R}^L$ is the latent body representation with dimensionality $L$, which has replaced $\mathbf{m}_\text{body}$ from the explicit representation. 
In practice, $\mathbf{x}_\text{body}$ is the output of a learned tokenizer (\cref{sec:tokenizer}) and each token encodes multiple frames of motion.
The diffusion model introduced in \cref{sec:diffusion} learns to generate motion using the hybrid representation, which has several advantages.
Maintaining root position features in global coordinates avoids potential compounding errors inherent to integrating local velocity-based representations.
The global root also facilitates controllable motion generation conditioned on spatial constraints, which are often sparse and defined within the global scene space, as it enables direct overwriting of root features. 
Moreover, the latent body representation is more compact than explicit representations, and pre-defined after the tokenizer is trained. This makes it better suited for generative modeling, both computationally and in terms of learning efficiency.

\subsection{Body Motion Tokenizer}
\label{sec:tokenizer}

\begin{figure}[t]
    \centering
    \includegraphics[width=\linewidth]{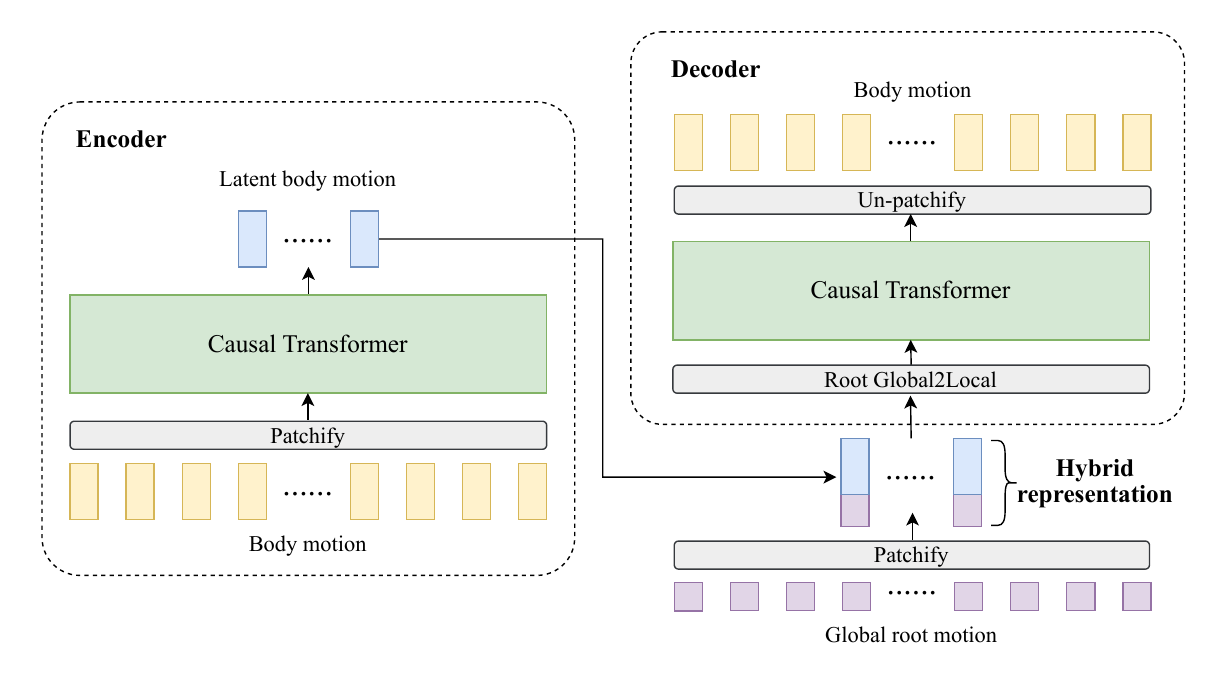}
    \caption{\textbf{Motion Tokenizer.} The encoder first embeds the patchified body motion into a latent representation. This latent body motion is concatenated with the patchified global root motion to form our \textit{hybrid representation}, which is decoded back to reconstruct the body motion. } 
\label{fig:tokenizer}
\end{figure}

We train a motion tokenization network to compress the high-dimensional explicit body features into a compact latent space, facilitating more efficient generative learning.
As illustrated in \cref{fig:tokenizer}, the tokenizer employs an asymmetric conditional autoencoder architecture.
Given an explicit body motion $\mathbf{m}_\text{body}^{1:N}$ containing $N$ frames, we treat each $P$ consecutive frames as a patch by reshaping them into a single vector, resulting in $T=N/P$ input vectors to the encoder.
The encoder compresses the body motion into latent tokens $\mathbf{x}_\text{body}^{1:T} \in \mathbb{R}^{T \times L}$, which are then concatenated along the feature dimension with the patchified explicit root motion $\mathbf{m}_\text{root}^{1:T} \in \mathbb{R}^{T\times 5P}$ to form the \textit{hybrid motion tokens}:
\begin{equation}
    \mathbf{x}^{1:T} = [ \mathbf{m}_\text{root}^{1:T}; \mathbf{x}_\text{body}^{1:T}  ]
\end{equation}
resulting in $\mathbf{x}^{1:T} \in \mathbb{R}^{T \times D}$ where $D = L + 5P$. 
The decoder subsequently reconstructs the body motion from these hybrid tokens.
Crucially, the decoder first transforms the global root motion from \cref{eqn:explicit_rep} into a local representation, which replaces the global root motion for the conditional input to the decoder network.
Each root pose in the local representation is a tuple $(\dot{\psi}, \dot{\mathbf{p}}_{x}, \dot{\mathbf{p}}_{z}, \mathbf{p}_y)$ where $\dot{\psi}$ is the 1D angular velocity of the heading, $\dot{\mathbf{p}}_{x}$ and $\dot{\mathbf{p}}_{z}$ are the x and z components of the linear root velocity, and $\mathbf{p}_y$ is the $y$-component (height) of the root.
Note that while the global root representation is useful for \textit{generating} motion as discussed previously, in the tokenizer decoder we find the local representation is more suitable to significantly mitigate foot skating (discussed in \cref{sec:results_ablation} and \cref{tab:ablation}).

We use transformer encoder layers with causal attention in both the encoder and decoder, which ensures that each frame embedding relies only on preceding frames and preserves temporal causality.
We experimented with different autoencoder variants for the tokenizer, including variational autoencoder (VAE) \cite{kingma2014auto} and finite scalar quantization (FSQ) \cite{mentzer2023finite} variants, as detailed in \cref{sec:results_hyperparams}. 
While all variants perform similarly, we find that FSQ demonstrates better stability in training, making it the default tokenizer choice. 
Training details can be found in \cref{sec:impl_details}. %

\begin{figure*}[t]
    \centering
    \includegraphics[width=\linewidth]{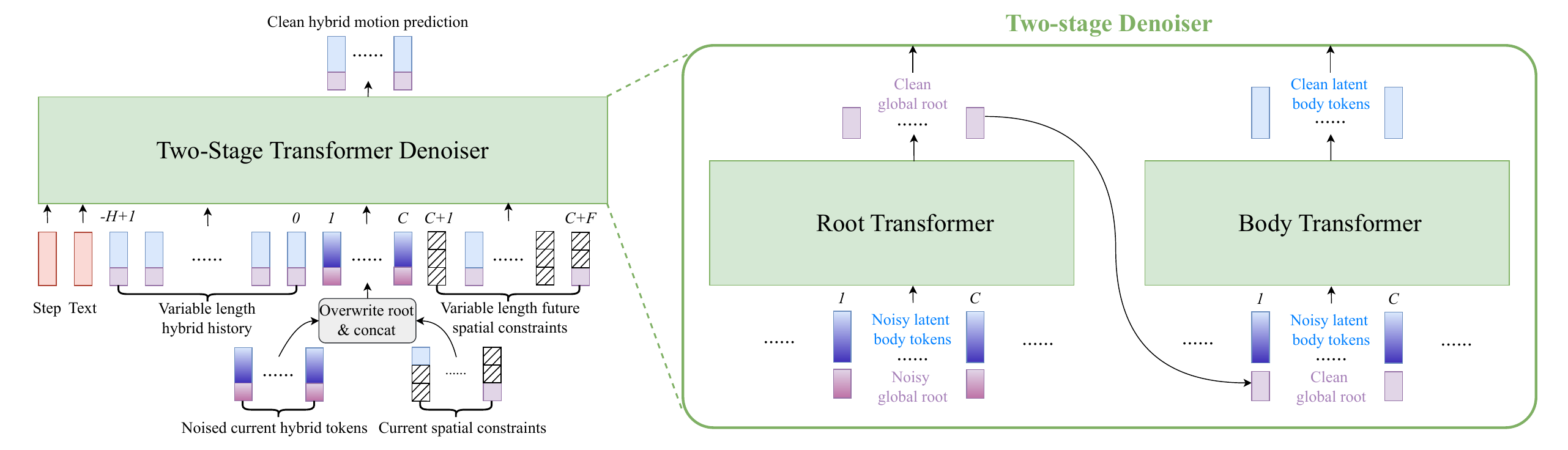}
    \caption{
    \textbf{Autoregressive Two-Stage Transformer Denoiser.} 
    (Left) Conditioned on a variable-length history context and optional spatial goal constraints, the autoregressive denoiser predicts a sequence of $C$ clean motion tokens within the current generation window. Spatial goal constraints can be arbitrarily sparse and may be located within or beyond the current motion generation window.
    (Right) The two-stage denoiser first predicts clean global root motion, which then conditions the second stage to predict clean latent body tokens, together forming the complete hybrid motion prediction.
} 
\label{fig:auto_transf}
\end{figure*}

\subsection{Controllable Interactive Motion Generation}
\label{sec:problem_formulation}
We aim to develop a motion generation model that supports text and spatial conditions from real-time input streams. 
At runtime, the model should be reactive to any changes in the input streams like a new text prompt or shift in goal location.
Similar to prior work~\cite{chen2024camdm,zhang2025primal,wu2025uniphys,tevet2025closd}, we formulate this problem as a conditional autoregressive generation task that synthesizes a short window of future motion starting from the current frame, conditioned on past history and optional goal inputs (\ie, kinematic constraints).
The synthesized future motion is then played back for the user until re-planning occurs and the model predicts future motion in the new window.

Our autoregressive model operates in the hybrid token space. 
Assuming that the prediction window starts at token index 1, then our goal is to train the generative model $\mathcal{F}$ to generate the next $C$ tokens in the \textit{current} prediction window:
\begin{equation}
\mathbf{x}^{1:C} = \mathcal{F}(s, \mathbf{x}^{(-H+1):0}, \mathbf{g}^{1:(C+F)}),
\label{equ:auto_chunk}
\end{equation}
where $s$ is the text prompt describing the motion semantics of the current generation window, $\mathbf{x}^{(-H+1):0}$ is the history motion spanning up to the previous $H$ tokens, and $\mathbf{g}^{1:(C+F)}$ denotes the spatial goals to achieve. Note that the goals for the first $C$ tokens $\mathbf{g}^{1:C}$ are within the current prediction horizon, while $\mathbf{g}^{(C+1):(C+F)}$ are goals beyond the prediction window, up to $F$ additional \textit{future} tokens. 

Notably, $H$ can vary in our formulation, so the model should expect to receive anywhere from 0 to a maximum of $H$ history conditioning tokens. 
A long history context is crucial to handle text prompts that describe complex non-cyclic motions. For instance, the prompt ``walk forward, then bend over and pick something up before continuing to walk'' has walking before and after the pick-up action. In autoregressive formulations with limited history context~\cite{wu2025uniphys,tevet2025closd,zhao2025DartControl}, a model conditioned only on recent walking frames cannot determine whether a preceding pick-up action has already occurred or still needs to be generated, leading to inaccurate generations with missing or duplicated actions.

While autoregressive generation maintains temporal causality (\ie, there is no dependence on future frames), it can still be conditioned on future goals $\mathbf{g}^{1:(C+F)}$. 
Spatial goals encompass constraints on the motion that specify joint position and/or rotation values at specific timesteps in the future.
These can be used to hit 2D waypoints or follow full paths on the ground, full-body pose keyframes, sparse end-effector position constraints, and more.
Importantly, our formulation is not limited to goals within the current prediction window, but is also conditioned on goals further in the future. 
Out-of-window goal constraints implicitly determine the motion generation within the current window, even though they do not directly apply to the immediate frames. For example, when a human needs to run to a location in 10 seconds, the destination goal will determine in which direction the human should start moving from the first step. 
Supporting such long-horizon goals in previous works requires training an additional RL control policy on top of the autoregressive motion model~\cite{shi2024amdm,zhao2025DartControl}, while our model architecture supports this natively.

\subsection{Autoregressive Two-Stage Diffusion Model}
\label{sec:diffusion}
Based on the hybrid motion representation, we design a transformer-based diffusion denoising model to  learn the goal-conditioned autoregressive motion generation task. To further enable precise controllability without sacrificing motion fidelity, we introduce an interleaved two-stage diffusion framework that decomposes the generation of root motion and body motion.

For an introduction to human motion diffusion, we refer the interested reader to prior work~\cite{tevet2023mdm,zhang2024motiondiffuse}, and focus here on relevant details for our method. 
At step $k$ in the denoising process, our diffusion model takes $C$ noisy hybrid motion tokens $\mathbf{x}^{1:C}_k \in \mathbb{R}^{C \times D}$ within the current generation window, along with relevant conditioning, and outputs a prediction for the clean denoised hybrid tokens $\hat{\mathbf{x}}^{1:C}_0$.
Mirroring \cref{equ:auto_chunk}, the denoising process at step $k$ can be written as:
\begin{equation}
    \hat{\mathbf{x}}^{1:C}_0 = \mathcal{F}(k, s, \mathbf{x}^{1:C}_k, \mathbf{x}^{(-H+1):0}, \mathbf{g}^{1:(C+F)}).
\label{eqn:denoiser}
\end{equation}
The high-level architecture of the denoising network is illustrated in the left side of~\cref{fig:auto_transf}.
The diffusion step and text conditioning are each a single token fed in alongside the sequence of history tokens and noisy tokens for the current prediction window. 
We use sinusoidal positional encodings for motion tokens to embed their temporal position within the motion sequence, while employing separate learned positional embeddings for text and diffusion tokens.
Linear layers are used to project all token types to the same feature dimensionality before feeding to the denoiser.

\paragraph{Spatial Goal Conditioning}
We represent spatial goal inputs $\mathbf{g}$ with a masked version of the explicit motion representation from \cref{eqn:explicit_rep}. This allows handling arbitrarily sparse global signals on any pose feature, such as keyframed body or end-effector joints. Only the constrained features and timesteps in $\mathbf{g}$ contain non-zero values while other unconstrained entries are set to zero.
We additionally define a corresponding binary mask $\mathbf{v}$ of the same shape, which indicates the dimensions that are constrained.
To align with the temporal granularity of the motion tokens, we assume the goal inputs are patchified, for example the short-term goals are $\mathbf{g}^{1:C} \in \mathbb{R}^{C \times MP}$ with patch size $P$ and pose feature dimensionality $M$.

Before being given to the model, the root part of the noisy tokens $\mathbf{m}^{1:C}_\text{root}$ is \textit{overwritten} with the root component of the constraint as $\tilde{\mathbf{m}}^{1:C}_\text{root} = (1-\mathbf{v}_\text{root}) \odot \mathbf{m}^{1:C}_\text{root} + \mathbf{v}_\text{root} \odot \mathbf{g}^{1:C}_\text{root}$ where $\odot$ is the elementwise product.
This root constraint overwriting~\cite{cohan2024condmdi, Kimodo2026} facilitates highly accurate control over the root trajectory, which governs the fundamental global movement of the human motion.
To incorporate constraints on detailed body poses and make the model aware of all constraints, we concatenate the explicit body goal features and the full constraint mask with the input tokens along the feature dimension.
In other words, the input noisy tokens are extended with masked constraints to form the augmented representation $[\tilde{\mathbf{m}}^{1:C}_\text{root}; \mathbf{x}^{1:C}_\text{body}; \mathbf{g}^{1:C}_\text{body}; \mathbf{v}]$ where $\mathbf{x}^{1:C}_\text{body}$ is the latent body part of the input noisy tokens.
Since there are no noisy input tokens beyond the prediction horizon $C$, the patchified long-horizon constraints $\mathbf{g}^{(C+1):(C+F)} \in \mathbb{R}^{F \times MP}$ are simply concatenated with their corresponding binary mask and fed in as additional tokens to the transformer. 
These long-horizon goal tokens can vary in length and sparsity depending on user input, with unconstrained tokens masked out during transformer inference.

\paragraph{Interleaved Two-Stage Denoiser}
Our autoregressive transformer denoiser employs an interleaved, two-stage design \cite{Kimodo2026} to sequentially predict clean root and body motions.
The internals of our transformer-based denoiser are shown on the right side of \cref{fig:auto_transf}.
At each denoising step, the model first predicts the explicit clean global root motion $\hat{\mathbf{m}}^{1:C}_{\text{root}}$ with the root transformer.
Next, the global root motion is detached and fed into the body transformer, which predicts the clean latent body tokens $\hat{\mathbf{x}}^{1:C}_{\text{body}}$. 
The outputs from both branches are concatenated to form the clean hybrid motion prediction $\hat{\mathbf{x}}^{1:C}_0 = [\hat{\mathbf{m}}^{1:C}_{\text{root}}; \hat{\mathbf{x}}^{1:C}_{\text{body}}]$. 
During inference, this concatenated hybrid prediction is re-noised for the subsequent diffusion step and fed back into the two-stage denoiser. This iterative and interleaved denoising process ensures continuous mutual influence between the root and body transformers throughout generation.
Finally, the predicted hybrid motion representation is processed by the tokenizer's decoder to recover the explicit body motion and form the full, un-patchified explicit motion as $\hat{\mathbf{m}}_0^{1:G} = [\hat{\mathbf{m}}^{1:G}_{\text{root}}; \hat{\mathbf{m}}^{1:G}_{\text{body}}]$, where the generation window size in frames is $G = C \cdot P$.

Our two-stage architecture is motivated by the hypothesis that predicting body motion conditioned on clean root motion is an easier task than generating both root and body jointly.
This decomposition is designed to enable precise controllability without compromising the fidelity of the synthesized motion. As demonstrated in our ablation study in \cref{tab:ablation}, the proposed two-stage architecture yields better results compared to a monolithic one-stage baseline that simultaneously predicts root and body motion.

\subsection{Training and Implementation Details}
\label{sec:impl_details}
\paragraph{Motion Tokenizer}
In practice, our motion tokenizer uses a patch size of $P = 4$ frames. Both the encoder and decoder are implemented as 8-layer transformers with a latent dimension of 512, utilizing causal self-attention to preserve temporal consistency.
The tokenizer is trained on motion clips of varying lengths (1–10 seconds) using a reconstruction loss and additional loss penalizing foot skating:
\begin{equation}
    \mathcal{L}_{\text{skate}} = \frac{\sum_{j \in \mathcal{S}_f} \hat{\mathbf{c}}_j \| \hat{\dot{\mathbf{J}}}_{j} \|_2}{\sum_{j \in \mathcal{S}_f}  \hat{\mathbf{c}}_j },
\end{equation}
where $\mathcal{S}_f$ represents the set of foot joint indices, $\hat{\mathbf{c}}_j$ denotes the predicted contact label for foot joint $j$, and $\| \hat{\dot{\mathbf{J}}}_{j} \|_2$ denotes the magnitude of predicted foot joint velocity. This foot-skating loss penalizes the velocities of joints predicted to be in contact with the ground, thereby enforcing stationary constraints during the contact phase. We set the weight for this loss term to 0.01.
The exact implementation of the reconstruction loss depends on the framework being employed for the tokenizer. We test three different approaches including a vanilla continuous autoencoder, VAE, and finite scalar quantization (FSQ)~\cite{mentzer2023finite} and compare them in experiments later (\cref{sec:results_hyperparams}). 
For the FSQ variant, we apply finite quantization to the encoder output embedding, constraining each feature to one of 64 discrete levels. These quantized vectors serve directly as the latent representation.
For all tokenizer variations, we train with the AdamAtan2~\cite{everett2024scaling} optimizer for 4 million steps using a learning rate of $2e{-}5$ and batch size of 128. We employ a cosine learning rate scheduler with a 10k-step linear warmup phase. Training is performed on a single NVIDIA A100-SXM4-80GB GPU. 

\paragraph{Two-Stage Denoiser} 
Both the root and body transformer in our two-stage denoiser employ the same transformer encoder architecture. Each transformer contains 8 layers with 8 heads and a latent size of 1024, totaling around 156 million parameters for our deployed denoiser model in the interactive demo. 
For text encoding, we use LLM2Vec~\cite{reddy2024llm2vec}, which is an embedding model trained on top of Llama-3-8B-Instruct~\cite{llama3modelcard}.

After training the tokenizer, we train the denoiser using the DDPM framework~\cite{ho2020denoising} with a modified version of the ``simplified'' loss function that contains several components. In the following discussion, we drop the token/frame index superscripts from all terms for simplicity.
First, given the clean hybrid prediction $\hat{\mathbf{x}}_0 = [\hat{\mathbf{m}}_{\text{root}}; \hat{\mathbf{x}}_{\text{body}}]$ and ground truth $\mathbf{x}_0$, the hybrid loss
\begin{align}
    \mathcal{L}_\text{hybrid} = || \hat{\mathbf{x}}_0 - \mathbf{x}_0 ||_1
\end{align}
uses a smooth L1 loss \cite{girshickFastRCNN2015} to penalize errors between the predicted and ground truth hybrid motion tokens. 
For the next loss, we decode the predicted tokens with the tokenizer decoder $\mathcal{D}$ resulting in the predicted explicit body motion $\hat{\mathbf{m}}_\text{body} = \mathcal{D}(\hat{\mathbf{x}}_0)$. Then, the decoded body loss 
\begin{align}
    \mathcal{L}_\text{dec} = || \hat{\mathbf{m}}_\text{body} - \mathbf{m}_\text{body} ||_1
\end{align}
compares the predicted explicit body motion to the ground truth $\mathbf{m}_\text{body}$. 
To place greater emphasis on accurately hitting the specified constraints, we add a goal loss
\begin{align}
    \mathcal{L}_\text{goal} = || \mathbf{v} \odot (\hat{\mathbf{m}}_0 - \mathbf{g}) ||_1
\end{align}
that specifically penalizes components in the full explicit motion prediction $\hat{\mathbf{m}}_0$ that do not hit the constraint goals in $\mathbf{g}$. 
Finally, we add a regularizer to ensure consistency between the directly predicted joint positions and those resulting from the predicted joint rotations via forward kinmeatics:
\begin{align}
    \mathcal{L}_\text{consist} = || \hat{\mathbf{J}}_0 - \text{FK}(\hat{\boldsymbol{\theta}}_0) ||_2
\end{align}
where $\hat{\mathbf{J}}_0$ denotes the predicted joint positions, and the forward kinematics function (FK) outputs joint positions given the predicted joint rotations $\hat{\boldsymbol{\theta}}_0$.
The final loss combines all these objectives as
\begin{align}
    \mathcal{L} = \mathcal{L}_\text{hybrid} + \mathcal{L}_\text{dec} + \mathcal{L}_\text{goal} + \mathcal{L}_\text{consist}.
\end{align}

The two-stage denoiser is trained on sequences with a maximum length of 10 seconds following existing offline motion generation works~\cite{tevet2023mdm, pinyoanuntapong2025MaskControl}. For each training motion sequence, a fixed-size generation window of $G$ frames is sampled randomly. Consequently, the lengths of the available history ($H$) and future ($F$) contexts for each training sample vary dynamically, ranging from 0 to the maximum sequence length minus $G$. 
Moreover, we augment the motion sequences by applying random rotations around the $y$-axis.
Spatial constraints for both in-horizon and out-of-horizon are randomly sampled from a set of common use cases including 2D root keyframes, 2D root trajectories, full-body sparse keyframes, full-body keyframe blocks, sparse end-effector keyframes, and foot contact keyframes.
To enable classifier-free guidance \cite{hoClassifierFreeDiffusionGuidance2022} during inference, we randomly drop the text prompts and spatial constraints with a 10\% probability. 

By default, we use ten diffusion steps during both train and test-time, which strikes a good balance between speed and accuracy. However, performance is still acceptable for most applications when going as low as four steps (see \cref{sec:results_hyperparams}). 
Denoiser training uses the AdamAtan2 optimizer with a learning rate of $2e{-}5$. 
Importantly, we do not use dropout in the denoiser as this causes root constraint conditioning inputs to be partially lost.
Our denoiser models are trained with a batch size of 512 across four NVIDIA A100-SXM4-80GB GPUs for one million optimization steps.

\begin{figure}[t]
    \centering
    \includegraphics[width=\linewidth]{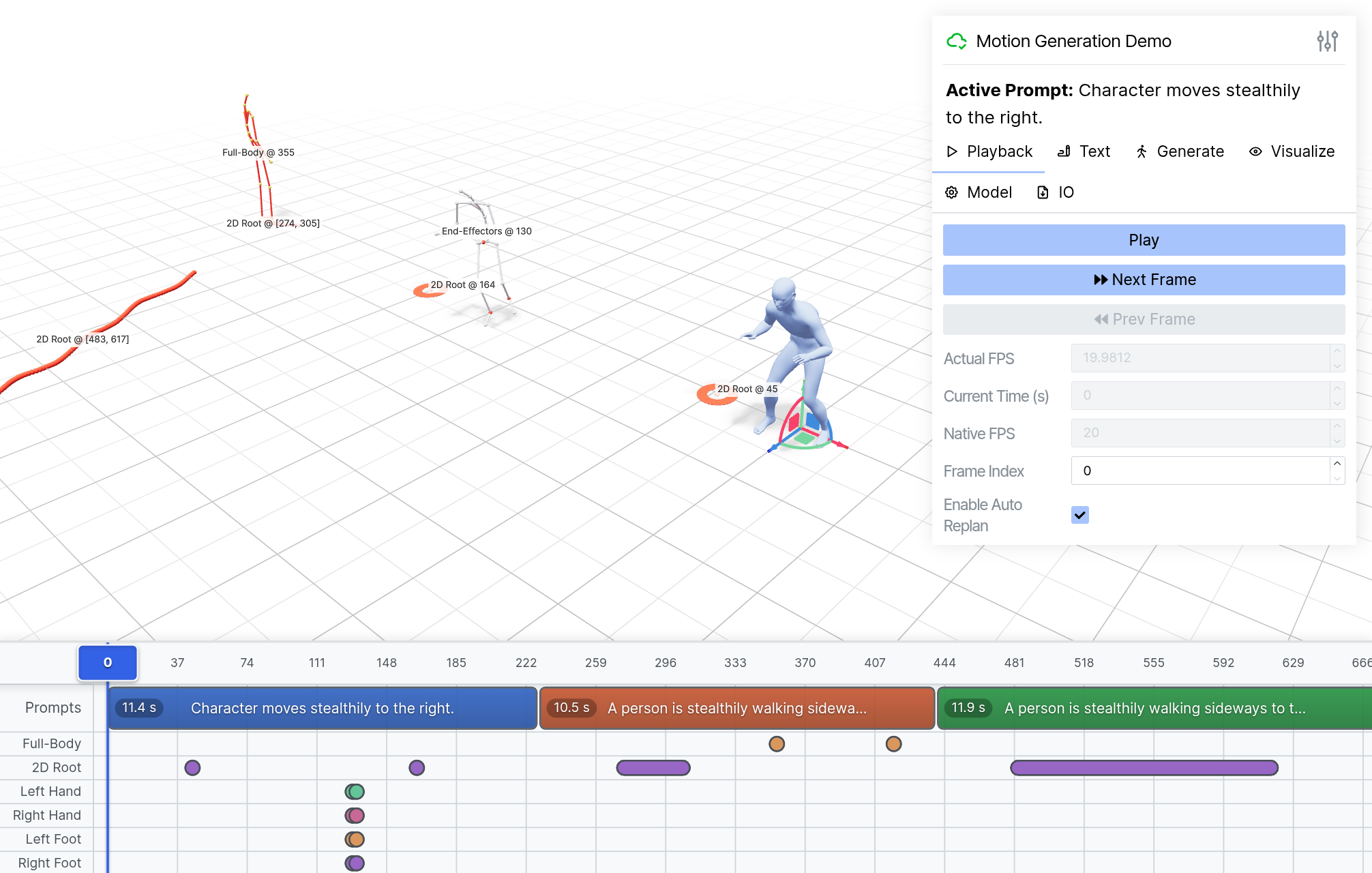}
    \caption{\textbf{Interactive Demo Interface.} This web interface allows generating motion and interacting with \method in real time. The \textit{control panel} at the top right allows dynamically changing the text prompt or input constraints. Input constraints are visualized in red within the \textit{3D scene} as the model generates motion to follow them. The \textit{timeline tracks} on the bottom of the interface intuitively show upcoming text prompts and constraints.} 
\label{fig:UI}
\end{figure}

\section{Interactive Motion Generation Demo}
\label{sec:results_demo}

To showcase \method's versatility, we developed an interface using Viser~\cite{yi2025viser} to interactively generate motion with our model. The system, shown in \cref{fig:UI}, enables real-time character control through a combination of streaming text prompts and interactive spatial constraints provided via mouse and keyboard inputs. 
In this section, we first detail \method's test-time operation, then qualitatively demonstrate key results through the interactive demo. 

\begin{figure}[t]
    \centering
    \includegraphics[trim={9mm 2mm 7mm 9mm}, clip,width=\linewidth]{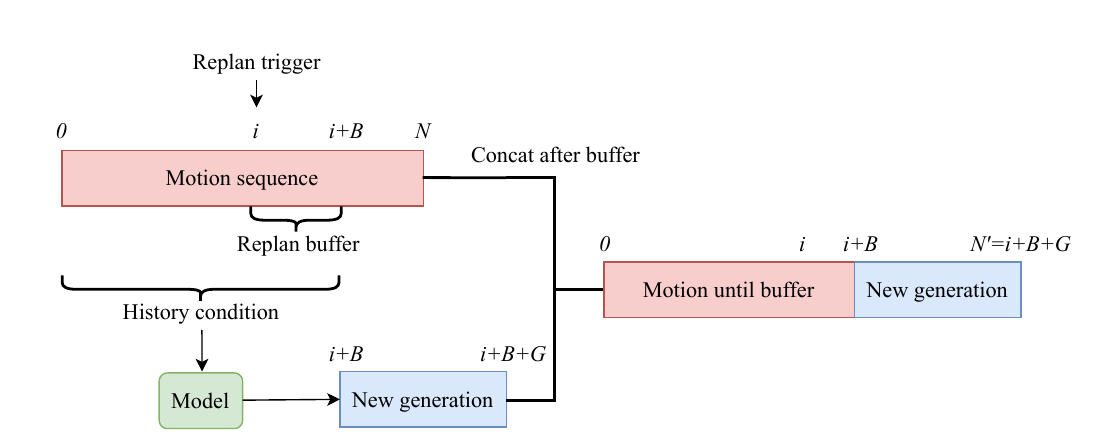}
    \caption{\textbf{Latency-Aware Replanning.} We utilize a non-blocking strategy where a buffer of $B$ frames is simultaneously played back and fed into the generation thread as history context. This buffer effectively hides the inference latency of slower models, ensuring that the transition to the newly generated sequence remains smooth and continuous. 
    } 
\label{fig:replan}
\end{figure}

\begin{figure*}[t]
    \centering
    \includegraphics[width=0.97\linewidth]{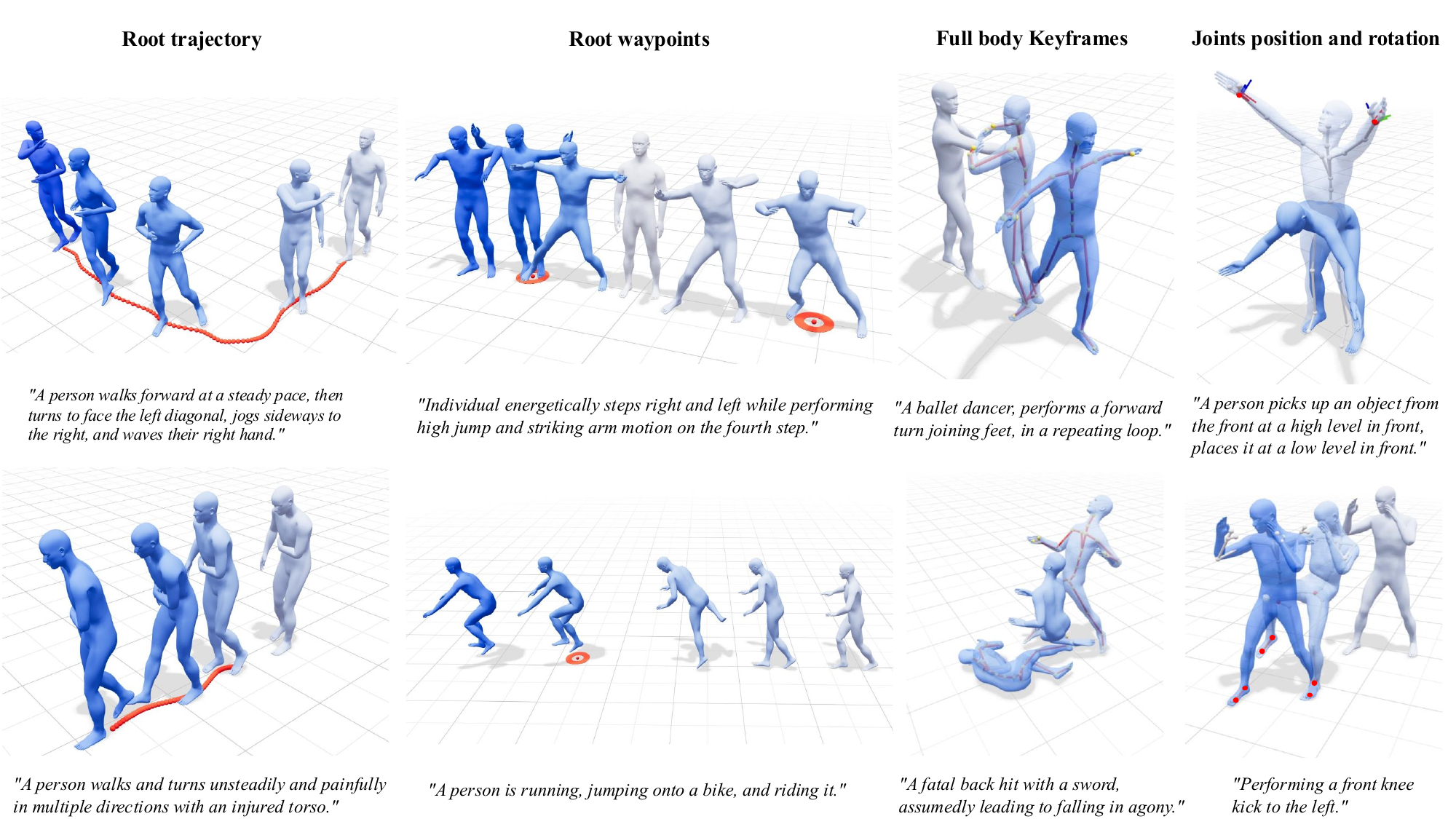}
    \caption{\textbf{Motion Generation with Kinematic Constraints.} 
    Qualitative results for motion generation conditioned on text prompts and diverse kinematic constraints, including dense root trajectories, sparse root waypoints (visualized as red rings), full-body keyframes (visualized as red skeletons), sparse joint positions (visualized as white skeletons with constrained joints highlighted as red spheres), and joint rotations (visualized as coordinate axes centered at the constrained joint). Motion temporal progression is indicated by a color gradient from gray to blue.
    } 
\label{fig:qualitative_results}
\end{figure*}
\subsection{Test-Time Operation}
During inference, \method operates autoregressively to synthesize motion in response to a dynamic stream of user inputs. 
In the first step of the motion roll-out, \method generates the first window of length $G$ with no history poses as input. In subsequent steps, the previously predicted tokens become the history conditioning as the model predicts the next window of $G$ motion frames. 
To facilitate autoregressive long motion generation, we employ a truncated sliding window to manage both historical and beyond-generation future contexts. The specific truncation lengths of these context windows are configurable in our interactive demo, up to a maximum of 8 seconds—a limit established by the longest context observed during training. Future constraints that fall beyond the truncation limit (e.g., a target location one minute ahead) are excluded from the input constraint tokens. They are only incorporated into the conditioning once the advancing generation window brings them within the truncated future context horizon.
During the autoregressive generation, the root component of the previously predicted tokens are translated such that the last frame of the history coincides with the origin, which is what the model expects as input. The translation offset is preserved and subsequently applied to the generated motion to transform it back into global scene coordinates. This loop ensures high-quality motion with smooth temporal transitions.

To enable real-time interactivity, we incorporate a dynamic replanning mechanism that triggers immediately upon detecting new user input, such as updated text prompts or modified future kinematic constraints, or when the current motion buffer will soon be depleted. 
Our replanning scheme is latency-aware, facilitating the use of more powerful models even when their inference latency exceeds the inter-frame interval. 
As shown in \cref{fig:replan}, when a replan is triggered we utilize the subsequent $B$ frames, which have already been generated, as a \textit{replan buffer}. These frames are played back to the user while simultaneously serving as history context for the asynchronous generation thread. This replan buffer effectively masks the inference latency of slower models, ensuring smooth and continuous transitions to the new generation. We present this scheme as an optional mechanism to enable increased diffusion steps for enhanced motion quality and control accuracy. In our deployment setup, the 4-step model operates without buffer frames, while the 10-step model employs a single buffer frame.

\subsection{Demo Results}
The interactive motion generation demo uses \method trained on the Bones Rigplay dataset~\cite{BonesStudio2026} described in detail later (see \cref{sec:results_internal}). The demo runs on a workstation equipped with an RTX 4090 GPU. The average generation latency is 33 ms for our efficient 4-step diffusion model and 63 ms for our 10-step diffusion model, with the latter providing slightly improved control accuracy.
Both models use a generation window of $G = 40$ frames (2 seconds at 20~fps).
All examples in \cref{fig:teaser} are generated using this interface, demonstrating that the system can process complex descriptions and seamlessly adapt to dynamic changes in user-specified text prompts. It also robustly satisfies diverse kinematic constraints, ranging from sparse long-term goals (e.g., reaching a target location in 10 seconds) to dense short-term constraints (e.g., trajectory following or full-body keyframes). 
Additional qualitative results for kinematic constraint-conditioned generation are shown in \cref{fig:qualitative_results}.

Our system also supports diverse locomotion interfaces: users can define target root trajectories in real time using mouse-based waypoints or modulate real-time velocity via keyboard commands.
For mouse-based root path control, we derive the target trajectory by linearly interpolating between mouse-click waypoints and smoothing the resulting path.
For keyboard-based root velocity control, we compute a target velocity from user input and the current velocity, then linearly interpolate between the two and integrate the resulting per-frame velocities to derive the root trajectory input to the model.
Extensive video demonstrations of our interactive generation system are provided in the supplementary material, highlighting its responsiveness and high-fidelity motion quality.

\section{Analysis on Large-Scale Mocap Data}
\label{sec:results_internal}
Next, we thoroughly analyze key design choices of \method along with the effects of various hyperparameter settings.

\subsection{Experiment Setting}
\paragraph{Bones Rigplay Mocap Dataset}
We leverage the large-scale proprietary Bones Rigplay dataset~\cite{BonesStudio2026}, which contains around 700 hours of diverse studio-quality human motion with text descriptions.
The scale and quality of this data enables a more robust testbed for evaluating design variations compared to smaller public datasets like HumanML3D~\cite{guo2022humanml3d}, which are saturated as indicated by methods scoring higher than ground truth data on metrics like R-precision.
This dataset contains motions from more than 150 participants and is retargeted to a unified-proportion 27-joint skeleton to facilitate learning. The motions encapsulate thousands of distinct behaviors, each performed by multiple actors for multiple takes, resulting in a diverse distribution of semantics and kinematic variations. It includes common motion categories such as locomotion, everyday activities, gestures, and combat, performed in a variety of styles. Raw motion clips range from 1 to 180 seconds in length, but we clip motions to a maximum of 10 seconds and subsample to 20 fps for training. 
To improve generalization, we use LLM to generate diverse paraphrases of the original text labels. 
The dataset is split into training and test sets by first grouping motion clips according to semantic content (\ie, action type, such as ``eating\_apple\_right''), and then assigning disjoint groups to each split with an approximate 90/10 ratio, resulting in about 315k motion clips for training and 35k for testing. As a result, the test set contains motion categories that are entirely unseen during training, providing a stronger evaluation of generalization to novel actions.

\paragraph{Constraints Sampling}
We evaluate text+constraint-conditioned generation across a comprehensive suite of test cases designed to simulate common downstream applications. These scenarios include dense root trajectory following, sparse waypoints navigation, full-body keyframes, and end-effector joints control (incorporating both position and orientation goals). The spatial constraints are sampled directly from the ground-truth test set alongside their corresponding text prompts. Furthermore, to rigorously evaluate the model's robustness against constraint inputs, we introduce slight random perturbations to the global translation and heading of a subset of sampled constraints during the evaluation.

\paragraph{Evaluation Metrics}
Following established protocols~\cite{guo2022humanml3d}, we employ Fréchet Inception Distance (\textbf{FID}) to quantify the distributional similarity between generated and ground-truth motions, and Top-3 \textbf{R-precision} to assess text-motion alignment. To ensure a rigorous evaluation, we train a robust evaluator model based on TMR~\cite{petrovich23tmr} using the large-scale Bones Rigplay dataset. Notably, we compute R-precision over a test dataset containing about 5k unique samples of diverse action types. This significantly increases the retrieval difficulty compared to the standard practice in benchmarks like HumanML3D~\cite{guo2022humanml3d}, which computes the metric over batches of size 32 only. As a proxy for motion quality, we also report a heuristic \textbf{foot skating} metric that measures mean foot velocity when the foot is considered in-contact based on a height threshold. To assess spatial control accuracy in constraints-conditioned generation, we compute the \textbf{mean error} between the user-specified constraint targets (position and orientation) and the corresponding generated poses.

\begin{table*}[t]
\centering
\caption{\textbf{Quantitative Ablation of Architectural Designs.} We evaluate performance across text-only and various kinematic constraints-conditioned generation scenarios, including end-effector joint rotation and position, full-body keyframe joints, dense root trajectories, and sparse root waypoints. $\uparrow$ denotes higher values are better; $\downarrow$ denotes lower values are better. \textbf{Bold} and \underline{underlined} values indicate the best and second-best results, respectively.}
\resizebox{\textwidth}{!}{
    \begin{tabular}{l ccc | cccccc}
    \toprule
    & \multicolumn{3}{c}{Text-only Generation} & \multicolumn{6}{c}{Constraints-conditioned Generation} \\
    Model & Skate (m/s) $\downarrow$ & R-prec. $\uparrow$ & FID $\downarrow$ & Skate (m/s) $\downarrow$ & Joint rot. (deg.) $\downarrow$ & Joint pos. (m) $\downarrow$ & Keyframe body (m)$\downarrow$ & Traj. (m) $\downarrow$ & Waypoint (m) $\downarrow$ \\
    \midrule
    Dataset & 0.255 & 76.56 & 0.000 & - & - & - & - & - & - \\
    \midrule
    \method (Ours) & \textbf{0.264} & \underline{65.47} & \textbf{0.027} & \underline{0.250} & \underline{2.23} & \textbf{0.025} & \textbf{0.023} & \textbf{0.015} & \textbf{0.024} \\
    Explicit representation & 0.365 & 53.90 & 0.065 & 0.281 & \textbf{1.67} & 0.130 & 0.136 & 0.033 & 0.203 \\
    Global root-conditioned decoder & 0.303 & 64.94 & \underline{0.028} & 0.284 & 2.88 & \underline{0.048} & \underline{0.044} & 0.024 & \underline{0.060} \\
    One-stage architecture & \textbf{0.264} & \textbf{65.84} & 0.029 & \textbf{0.248} & 2.46 & 0.101 & 0.079 & \underline{0.017} & 0.164 \\
    \bottomrule
    \end{tabular}
}
\label{tab:ablation}
\end{table*}

\begin{table*}[h]
\centering
\caption{\textbf{Hyperparameter and Tokenizer Analysis.} The \textbf{best} results in each group are highlighted in bold, and the \underline{second best} are underlined. The ablation table is divided into five sections, sequentially comparing: (1) generation horizon, (2) diffusion steps, (3) tokenizer patch sizes, (4) tokenizer latent space capacities (latent embedding quantization levels and dimensions), and (5) various tokenizer types. The default configuration in each section is marked with $^*$.
}
\resizebox{\textwidth}{!}{
    \begin{tabular}{l cc | cccc cccccc}
    \toprule
    \multicolumn{3}{c|}{Model} & \multicolumn{3}{c}{Text-only Generation} & \multicolumn{6}{c}{Constraints-conditioned Generation} \\
    \cmidrule(r){1-3} \cmidrule(lr){4-6} \cmidrule(l){7-12}
    Tokenizer & Horizon & Diffusion step & Skate (m/s) $\downarrow$ & R-prec. $\uparrow$ & FID $\downarrow$ & Skate (m/s) $\downarrow$ & Joint rot. (deg.) $\downarrow$ & Joint pos. (m) $\downarrow$ & Keyframe body (m)$\downarrow$ & Traj. (m) $\downarrow$ & Waypoint (m) $\downarrow$ \\
    \midrule
    \multicolumn{12}{c}{\textit{Generation horizon}} \\
    \midrule
    FSQ 64-128, Patch 4 & 4 & 10 & \textbf{0.151} & 33.42 & 0.224 & 0.445 & 9.23 & 0.848 & 0.864 & 0.864 & 0.850 \\
    FSQ 64-128, Patch 4 & 8 & 10 & 0.258 & 56.70 & 0.037 & \textbf{0.243} & 3.45 & \underline{0.031} & \underline{0.026} & \textbf{0.013} & \textbf{0.020} \\
    FSQ 64-128, Patch 4 & 12 & 10 & \underline{0.254} & 59.54 & 0.033 & \underline{0.247} & 2.94 & 0.033 & 0.028 & 0.017 & 0.031 \\
    FSQ 64-128, Patch 4 & 20 & 10 & 0.255 & \underline{63.80} & \underline{0.030} & 0.250 & \underline{2.61} & 0.046 & 0.037 & \underline{0.014} & 0.059 \\
    FSQ 64-128, Patch 4 & 40$^*$ & 10 & 0.264 & \textbf{65.47} & \textbf{0.027} & 0.250 & \textbf{2.23} & \textbf{0.025} & \textbf{0.023} & 0.015 & \underline{0.024} \\
    \midrule
    \multicolumn{12}{c}{\textit{Diffusion steps}} \\
    \midrule
    FSQ 64-128, Patch 4 & 40 & 1 & 0.411 & 56.74 & 0.079 & 1.405 & 25.39 & 1.040 & 1.054 & 1.037 & 1.002 \\
    FSQ 64-128, Patch 4 & 40 & 2 & 0.239 & 61.28 & 0.052 & 0.360 & 7.96 & 0.174 & 0.169 & 0.274 & 0.163 \\
    FSQ 64-128, Patch 4 & 40 & 3 & \underline{0.231} & 63.59 & 0.041 & 0.254 & 3.58 & 0.053 & 0.051 & 0.046 & 0.044 \\
    FSQ 64-128, Patch 4 & 40 & 4 & \textbf{0.230} & 64.41 & 0.034 & \textbf{0.249} & \underline{2.68} & 0.034 & 0.032 & 0.028 & \underline{0.027} \\
    FSQ 64-128, Patch 4 & 40 & 10$^*$ & 0.264 & \underline{65.47} & \underline{0.027} & \underline{0.250} & \textbf{2.23} & \textbf{0.025} & \textbf{0.023} & \underline{0.015} & \textbf{0.024} \\
    FSQ 64-128, Patch 4 & 40 & 100 & 0.282 & \textbf{65.49} & \textbf{0.025} & 0.257 & 2.71 & \underline{0.030} & \underline{0.027} & \textbf{0.009} & 0.028 \\
    \midrule
    \multicolumn{12}{c}{\textit{Tokenizer patch size}} \\
    \midrule
    FSQ 64-128, Patch 1 & 40 & 10 & \underline{0.298} & 44.45 & 0.152 & 0.355 & \underline{2.31} & 0.764 & 0.816 & 0.790 & 0.775 \\
    FSQ 64-128, Patch 4$^*$ & 40 & 10 & \textbf{0.264} & \underline{65.47} & \underline{0.027} & \textbf{0.250} & \textbf{2.23} & \textbf{0.025} & \textbf{0.023} & \textbf{0.015} & \textbf{0.024} \\
    FSQ 64-128, Patch 8 & 40 & 10 & 0.317 & \textbf{68.01} & \textbf{0.022} & \underline{0.295} & 3.05 & \underline{0.070} & \underline{0.062} & \underline{0.018} & \underline{0.100} \\
    \midrule
    \multicolumn{12}{c}{\textit{Tokenizer latent space capacity}} \\
    \midrule
    FSQ 16-32, Patch 4 & 40 & 10 & 0.283 & \textbf{68.11} & \textbf{0.023} & 0.261 & 4.57 & 0.031 & 0.026 & 0.016 & \underline{0.020} \\
    FSQ 64-32, Patch 4 & 40 & 10 & 0.273 & \underline{67.62} & \underline{0.025} & \underline{0.252} & 3.96 & \underline{0.026} & \textbf{0.023} & \textbf{0.014} & \textbf{0.017} \\
    FSQ 64-128, Patch 4$^*$ & 40 & 10 & \textbf{0.264} & 65.47 & 0.027 & \textbf{0.250} & \textbf{2.23} & \textbf{0.025} & \textbf{0.023} & \underline{0.015} & 0.024 \\
    FSQ 64-256, Patch 4 & 40 & 10 & \underline{0.268} & 64.04 & 0.031 & 0.257 & \underline{2.31} & 0.030 & 0.025 & \underline{0.015} & 0.032 \\
    \midrule
    \multicolumn{12}{c}{\textit{Tokenizer type}} \\
    \midrule
    AE 128D, Patch 4 & 20 & 10 & 0.266 & 62.20 & 0.033 & \textbf{0.246} & \underline{2.23} & \textbf{0.044} & \underline{0.040} & 0.016 & \textbf{0.057} \\
    VAE 128D, Patch 4 & 20 & 10 & \underline{0.259} & \underline{63.35} & \underline{0.031} & \underline{0.250} & \textbf{2.17} & \underline{0.046} & 0.042 & \textbf{0.014} & \underline{0.058} \\
    FSQ 64-128, Patch 4$^*$ & 20 & 10 & \textbf{0.255} & \textbf{63.80} & \textbf{0.030} & \underline{0.250} & 2.61 & 
    \underline{0.046} & \textbf{0.037} & \textbf{0.014} & 0.059 \\
    \bottomrule
    \end{tabular}
}
\label{tab:ablation_detailed}
\end{table*}

\subsection{Ablation Study}
\label{sec:results_ablation}

\cref{tab:ablation} presents ablation results on three key design choices: the hybrid motion representation, the global-to-local root conversion within the tokenizer decoder, and the two-stage denoiser design.

\paragraph{Hybrid Motion Representation}
We first compare our proposed hybrid motion representation (derived via the learned tokenizer) against the purely explicit motion representation. To ensure a fair comparison, we train an autoregressive baseline that uses explicit pose features, applying the same patching strategy to align the temporal granularity of the tokens. This explicit baseline uses masked overwriting (\cref{sec:diffusion}) to condition on all kinematic constraint inputs by overwriting both constrained root and body features.
As demonstrated in \cref{tab:ablation}, our autoregressive model utilizing the hybrid representation significantly outperforms its explicit counterpart in both motion quality and control accuracy. 
The high-dimensionality of explicit motion representations likely complicates the generative learning process, particularly under our few-step denoising setting. In contrast, the hybrid representation compresses high-dimensional body features into compact latent embeddings that are more amenable to efficient generative modeling.

\paragraph{Global-to-Local Conversion}
Next, we evaluate the importance of our global-to-local root conversion process within the tokenizer decoder by training a baseline decoder that operates directly on the global root representation. The ablation results reveal that removing the global-to-local root conversion leads to a notable increase in foot skating, confirming that local root representations are essential for preserving motion quality and physical plausibility.

\paragraph{Two-Stage Denoiser Design}
To validate our two-stage model architecture, we train a one-stage baseline that jointly predicts the root trajectory and latent body motion tokens simultaneously. Experimental results show that the two-stage architecture achieves superior performance, yielding higher-fidelity text-conditioned motion and significantly lower spatial constraint errors. This suggests that decomposing root and body prediction facilitates the simultaneous learning of high-fidelity generation and precise spatial control.

\subsection{Hyperparameter and Tokenizer Type Analysis}
\label{sec:results_hyperparams}

\cref{tab:ablation_detailed} provides an analysis of the generation horizon length, the number of diffusion steps, and the tokenizer configurations.

\paragraph{Generation Horizon}
The generation horizon length is a critical hyperparameter impacting the model's performance. We observe that extending the horizon consistently improves motion fidelity (FID) and semantic alignment (R-Precision) metrics. Conversely, extremely narrow horizons (\eg, 4 frames) lead to training instability and degraded performance, ultimately resulting in the generation of drifting motions. The text-only foot-skating metric for the 4-frame horizon is misleadingly low, as the model often fails to respond to text prompts. Regarding spatial control, we find that horizons of 8 and 40 frames effectively minimize the constraint errors.
Qualitative analysis shows that models with an 8-frame horizon can transition between actions more rapidly in response to updated text prompts compared to those with a 40-frame horizon. Furthermore, our experiments show that the 8-frame model learns constraint adherence faster during training than its 40-frame counterpart.

\paragraph{Diffusion Step}
We ablate the impact of the number of diffusion steps used by the autoregressive denoiser.
Using extremely few diffusion steps (\eg, 1 or 2) leads to significantly worse generation quality and constraint adherence.
Increasing diffusion steps provides slight gains in FID, R-Precision, and constraint accuracy. However, our few-step models still achieve highly competitive performance, demonstrating the robustness of the learned hybrid representation for efficient high-quality motion synthesis.

\paragraph{Tokenizer Patch Size}
We also evaluate the effect of the tokenizer patch size. Using a minimum patch size of a single frame leads to faster learning in the early stages, but causes training instability later on, resulting in significantly worse overall performance in the end. Conversely, using a larger patch size of 8 slightly improves the FID and R-precision metrics, but at the cost of worse skating performance and constraint accuracy. This trade-off occurs because compressing more frames into a single token causes a greater loss of fine-grained pose details within each patch.
\paragraph{Tokenizer Latent Space Capacity}
We evaluate tokenizers with varying latent space capacities. The capacity of a Finite Scalar Quantization (FSQ) latent space is determined jointly by the number of discrete quantization levels and the number of latent dimensions. By default, we use an FSQ configuration with 64 quantization levels and 128 dimensions, denoted as FSQ 64-128. While performance is relatively similar across configurations, there are some notable differences. Using FSQ 16-32 with a smaller latent capacity yields slightly better FID and R-precision metrics under the limited training budget of 1 million iterations, but it degrades performance on end-effector joint constraints and full-body errors. This trade-off arises because a smaller latent space lacks the capacity to represent fine-grained motion details accurately. On the other hand, expanding the number of dimensions to 256 slows model convergence and does not provide performance gains within the same train budget.

\paragraph{Tokenizer Type}
We experiment with several tokenizer architectures, including Variational Autoencoders (VAE) and Finite Scalar Quantization (FSQ). For the VAE variant, we applied a KL-divergence loss with weight of $1 \times 10^{-6}$ to regularize the latent distribution. Our results indicate that all tokenizer variants perform comparably to a vanilla autoencoder. 
However, the vanilla autoencoder suffers from severe training instability and diverges when trained with longer horizons, such as 40 frames. In contrast, the FSQ tokenizer demonstrates superior training stability over the vanilla autoencoder baseline, leading us to adopt FSQ as our default choice.

\section{Benchmark Evaluation}
\label{sec:results_compare}

Lastly, we evaluate \method against both offline and online state-of-the-art baselines for text+constraints-conditioned generation on the standard HumanML3D~\cite{guo2022humanml3d} dataset.
For these experiments, our model is trained with a 40-frame generation horizon using 10 diffusion steps and a vanilla autoencoder tokenizer.

\subsection{Experiment Setting}
\paragraph{HumanML3D Dataset}
This public dataset contains around 30 hours of motion data with corresponding text descriptions. In our experiments, we exclude the HumanAct12~\cite{guo2020action2motion} subset of HumanML3D due to the absence of native joint rotation data and the severe motion artifacts introduced by the original preprocessing.
During data processing, we preserve the original SMPL~\cite{SMPL:2015} joint rotations in the retargeting step, unlike the original HumanML3D pipeline, which discards native joint rotations. This makes our processed data compatible with real-time animation, since we can directly animate the body model with generated joint rotations instead of going through an expensive inverse kinematics post-process using generated joint positions.

\paragraph{Evaluation Metrics}
We adopt the evaluation benchmark from prior work~\cite{guo2022humanml3d,pinyoanuntapong2025MaskControl} to assess various aspects of the generated motion. To evaluate text-following, we report the Top-3 \textbf{R-precision}. Motion quality is measured via Fréchet Inception Distance (\textbf{FID}), which indicates similarity to the ground-truth distribution, and the \textbf{foot skating ratio}, which quantifies the frequency of detected foot skating frames. To assess spatial control accuracy, we calculate the \textbf{mean joint error} for the constrained goal joint positions.
We utilize the original HumanML3D evaluator models, which were trained on the original processed HumanML3D data, to calculate FID and R-precision metrics. As a result, our method is slightly disadvantaged on these metrics.
Additionally, we report the motion generation \textbf{latency} for each method, measured on a single NVIDIA A100-SXM4-80GB GPU.

\begin{table}[t]
\centering
\caption{\textbf{Offline Text and Constraint Control Comparison.} Evaluation results of text-conditioned motion generation with joint position goals on HumanML3D. $^*$ denotes methods without test-time optimization. $\uparrow$ denotes higher values are better; $\downarrow$ denotes lower values are better.}
\label{table:offline_compare}
\resizebox{\columnwidth}{!}{%
\begin{tabular}{lccccc}
\toprule
\textbf{Method} & \textbf{R-Prec. $\uparrow$} & \textbf{FID $\downarrow$} & \textbf{Skate (\%) $\downarrow$} & \textbf{Error (cm) $\downarrow$} & \textbf{Latency (s) $\downarrow$} \\
\midrule
Dataset (HumanML3D retarget) & 0.739 & 0.000 & 7.92 & 0.00 & - \\
Dataset (Our retarget)  & 0.732 & 0.011 & 6.87 & 0.00 & - \\
\midrule
\multicolumn{6}{c}{\textit{Without optimization}} \\
\midrule
MaskControl$^*$ \cite{pinyoanuntapong2025MaskControl} & \textbf{0.760} & 0.050 & 7.27 & 46.18 & 0.46 \\
\textbf{\method (Ours)}$^*$ & 0.729 & \textbf{0.044} & \textbf{6.28} & \textbf{4.15} & \textbf{0.15} \\
\midrule
\multicolumn{6}{c}{\textit{With optimization}} \\
\midrule
MaskControl \cite{pinyoanuntapong2025MaskControl} & \textbf{0.758} & \textbf{0.047} & 7.87 & 0.45 & 68.65 \\
\textbf{\method (Ours) Opt} & 0.721 & 0.088 & \textbf{5.87} & \textbf{0.30} & \textbf{9.25} \\
\bottomrule
\end{tabular}%
}
\end{table}

\subsection{Offline Model Comparison}

We first compare to MaskControl~\cite{pinyoanuntapong2025MaskControl}, a SOTA offline motion generation model that specializes in accurate joint controls. 
Following the protocol in MaskControl, we evaluate the model's ability to satisfy arbitrary joint position constraints at any given frame.
We first compare our raw generation results against MaskControl with its test-time optimization module disabled (denoted as MaskControl\textsuperscript{*}). Subsequently, we apply a similar test-time optimization to our predicted hybrid motion to minimize joint errors. We then compare these refined results against the full MaskControl pipeline. As shown in \cref{table:offline_compare}, \method achieves competitive text-following (on par with ground truth R-prec) and motion quality while demonstrating a lower foot skating ratio. Notably, compared to the raw MaskControl output before optimization, our method yields significantly lower spatial control errors, indicating a stronger underlying generative prior.

\subsection{Autoregressive Model Comparison}
Next, we compare \method to the closely related model DiP~\cite{tevet2025closd}, an autoregressive motion diffusion model. 
For the autoregressive model comparison, we evaluate constraints satisfaction by sampling goal joints using two distinct schemes. The first scheme, termed \textit{in-horizon goals}, follows the original DiP setting by sampling one goal joint at the final frame of each autoregressive generation window. This scheme necessitates a goal input every 2 seconds, which is often impractical for applications relying on sparser control signals. The second scheme, \textit{out-of-horizon goals}, involves sampling a single final goal joint at the very end of a long sequence which is beyond the initial autoregressive generation window. This configuration creates a challenging scenario requiring long-horizon planning, a task that the DiP system fails to handle effectively. 
Following the implementation of DiP, we sample the goal joints from the pelvis, wrists, and feet. 
We set the test sequence length to 9 seconds and provide 1 second of ground truth motion as initial history to adapt to the original DiP implementation.

As presented in \cref{table:autoregressive_compare}, our approach surpasses DiP in both in-horizon and out-of-horizon scenarios. Notably, DiP exhibits a sharp increase in joint error under the out-of-horizon setting, indicating its limitation for long-term planning. In contrast, our method effectively resolves these long-context constraints, maintaining high accuracy even when goals are placed far into the future.
Furthermore, to ensure our quantitative gains translate to actual human perception, we conduct a side-by-side perceptual study comparing the two methods on motion quality, semantic alignment, and joint goal accuracy for out-of-horizon goals. Participants are instructed to vote for the better result or indicate a tie. Across 240 pairwise human comparisons (\cref{tab:perceptual_study}), our approach \method is strongly and consistently preferred over DiP, confirming that the numerical improvements in \cref{table:autoregressive_compare} reflect genuine qualitative gains.

\begin{table}[t]
\centering
\caption{\textbf{Autoregressive Text and Constraint Control Comparison.} evaluation results of text-conditioned autoregressive motion generation with in-horizon and out-of-horizon sparse joint goals on HumanML3D. $\uparrow$ denotes higher values are better; $\downarrow$ denotes lower values are better.}
\label{table:autoregressive_compare}
\resizebox{\columnwidth}{!}{%
\begin{tabular}{lccccc}
\toprule
\textbf{Method} & \textbf{R-Prec. $\uparrow$} & \textbf{FID $\downarrow$} & \textbf{Skate (\%) $\downarrow$} & \textbf{Error (cm) $\downarrow$} & \textbf{Latency (s) $\downarrow$} \\
\midrule
Dataset (HumanML3D retarget) & 0.711 & 0.000 & 8.53 & 0.00 & - \\
Dataset (Our retarget) & 0.711 & 0.010 & 7.00 & 0.00 & - \\
\midrule
\multicolumn{6}{c}{\textit{In-horizon goals}} \\
\midrule
DiP \cite{tevet2025closd} & 0.609 & 0.967 & 12.29 & 9.20 & 0.15 \\
\textbf{\method (Ours)} & \textbf{0.690} & \textbf{0.092} & \textbf{7.07} & \textbf{2.48} & 0.15 \\
\midrule
\multicolumn{6}{c}{\textit{Out-of-horizon goals}} \\
\midrule
DiP \cite{tevet2025closd} & 0.599 & 1.453 & 11.07 & 17.64 & 0.15 \\
\textbf{\method (Ours)} & \textbf{0.684} & \textbf{0.100} & \textbf{7.63} & \textbf{2.92} & 0.15 \\
\bottomrule
\end{tabular}%
}
\end{table}

\section{Discussion}
\label{sec:discussion}
We propose \method, an autoregressive motion diffusion model that enables interactive and controllable human motion generation. \method natively supports online text prompting and flexible kinematic goal constraints tailored to interactive applications, including long-horizon goals that extend beyond a single generation window. 
We present a real-time demonstration of interactive and instructable motion generation, underscoring the potential of generative models for future animation systems.
We validate our architectural decisions through extensive ablation studies on the large-scale, studio-quality Bones Rigplay dataset. Furthermore, experiments on the public HumanML3D benchmark demonstrate that \method outperforms existing methods in terms of both motion fidelity and control accuracy.

\paragraph{Limitations}
While \method demonstrates a promising system for interactive human motion generation, several aspects of the design remain open for future research improvement.
First, \method explicitly utilizes all past motion frames as the history context during autoregressive generation, which can be inefficient for extremely long-horizon tasks. Exploring more efficient, structured memory representations and update mechanisms is an important future direction.
Second, as a diffusion model, \method relies on a multi-step iterative generation process, which can be computationally demanding. This could potentially be further accelerated by combining our approach with recent advances in shortcut diffusion models~\cite{lu2025simplifying, geng2025mean}.
Third, \method is a purely kinematic model and lacks awareness of physical dynamics. Consequently, artifacts such as foot skating and jittering can sometimes be observed in the generated motions. A crucial future direction is to integrate physics modelling into \method, proposing a unified generative model capable of predicting both the kinematics and dynamics of human motion, which is essential for physics-critical applications.

\begin{table}[t]
  \footnotesize
  \centering
  \caption{
  \textbf{Perceptual Study Results.} We report the percentage of human preferences comparing our method against DiP across three criteria. Our approach is consistently preferred over DiP, with a significant margin in motion quality, semantic alignment, and goal accuracy.
  }
  \label{tab:perceptual_study}
  \begin{tabular}{lccc}
    \toprule
    & \textbf{Ours (\%)} & \textbf{Tie (\%)} & \textbf{DiP \cite{tevet2025closd} (\%)} \\
    \midrule
    Motion Quality      & 65.8 & 25.0 & 9.2 \\
    Semantic Alignment  & 67.5 & 25.0 & 7.5 \\
    Goal Accuracy       & 64.6 & 31.2 & 4.2 \\
    \bottomrule
  \end{tabular}
\end{table}

\section{Acknowledgments}
We would like to thank Edy Lim, Eugene Jeong, Sam Wu, Ehsan Hassani, Michael Huang, and Jin-Bey Yu for their help with data processing and cleaning, and Cyrus Hogg, Simon Yuen, Lindsey Pavao, Jenna Diamond, Rizwan Khan, Samantha Shinagawa, and Akanksha Shukla for their efforts on data acquisition and labeling. We also thank the anonymous reviewers for their valuable feedback.

\bibliographystyle{ACM-Reference-Format}
\bibliography{main}

@String(CVPR= {IEEE Conf. Comput. Vis. Pattern Recog.})

@String(ICCV= {Int. Conf. Comput. Vis.})

@String(ECCV= {Eur. Conf. Comput. Vis.})

@String(TOG= {ACM Trans. Graph.})

@String(ICLR = {Int. Conf. Learn. Represent.})

@String(CVPR  = {CVPR})

@String(ICCV  = {ICCV})

@String(ECCV  = {ECCV})

@String(TOG   = {ACM TOG})

@String(ICLR  = {ICLR})

@inproceedings{
    tevet2023mdm,
    title={Human Motion Diffusion Model},
    author={Guy Tevet and Sigal Raab and Brian Gordon and Yoni Shafir and Daniel Cohen-or and Amit Haim Bermano},
    booktitle={The Eleventh International Conference on Learning Representations },
    year={2023},
    url={https://openreview.net/forum?id=SJ1kSyO2jwu}
}

@inproceedings{pinyoanuntapong2025MaskControl,
    title     = {MaskControl: Spatio-Temporal Control for Masked Motion Synthesis},
    author    = {Pinyoanuntapong, Ekkasit and Saleem, Muhammad and Karunratanakul, Korrawe and Wang, Pu and Xue, Hongfei and Chen, Chen and Guo, Chuan and Cao, Junli and Ren, Jian and Tulyakov, Sergey},
    booktitle = {Proceedings of the IEEE/CVF International Conference on Computer Vision (ICCV)},
    pages     = {9955--9965},
    year      = {2025}
}

@inproceedings{guo2024momask,
  title={Momask: Generative masked modeling of 3d human motions},
  author={Guo, Chuan and Mu, Yuxuan and Javed, Muhammad Gohar and Wang, Sen and Cheng, Li},
  booktitle={Proceedings of the IEEE/CVF Conference on Computer Vision and Pattern Recognition},
  pages={1900--1910},
  year={2024}
}

@inproceedings{dai2025motionlcm,
  title={Motionlcm: Real-time controllable motion generation via latent consistency model},
  author={Dai, Wenxun and Chen, Ling-Hao and Wang, Jingbo and Liu, Jinpeng and Dai, Bo and Tang, Yansong},
  booktitle={ECCV},
  pages={390--408},
  year={2025}
}

@inproceedings{
    tevet2025closd,
    title={{CL}o{SD}: Closing the Loop between Simulation and Diffusion for multi-task character control},
    author={Guy Tevet and Sigal Raab and Setareh Cohan and Daniele Reda and Zhengyi Luo and Xue Bin Peng and Amit Haim Bermano and Michiel van de Panne},
    booktitle={The Thirteenth International Conference on Learning Representations},
    year={2025},
    url={https://openreview.net/forum?id=pZISppZSTv}
}

@inproceedings{zhao2025DartControl,
   title = {{DartControl}: A Diffusion-Based Autoregressive Motion Model for Real-Time Text-Driven Motion Control},
   author = {Zhao, Kaifeng and Li, Gen and Tang, Siyu},
   booktitle = {The Thirteenth International Conference on Learning Representations (ICLR)},
   year = {2025}
}

@article{
        shi2024amdm,
        author = {Shi, Yi and Wang, Jingbo and Jiang, Xuekun and Lin, Bingkun and Dai, Bo and Peng, Xue Bin},
        title = {Interactive Character Control with Auto-Regressive Motion Diffusion Models},
        year = {2024},
        issue_date = {August 2024},
        publisher = {Association for Computing Machinery},
        address = {New York, NY, USA},
        volume = {43},
        journal = {ACM Trans. Graph.},
        month = {jul}
      }

@inproceedings{chen2024camdm,
  title={Taming Diffusion Probabilistic Models for Character Control},
  author = {Chen, Rui and Shi, Mingyi and Huang, Shaoli and Tan, Ping and Komura, Taku and Chen, Xuelin},
  year = {2024},
  publisher = {Association for Computing Machinery},
  address = {New York, NY, USA},
  url = {https://doi.org/10.1145/3641519.3657440},
  doi = {10.1145/3641519.3657440},
  booktitle = {ACM SIGGRAPH 2024 Conference Papers},
  location = {Denver, CO, USA},
  series = {SIGGRAPH '24}
}

@article{tessler2024maskedmimic,
    author = {
        Tessler, Chen and Guo, Yunrong and Nabati, Ofir and Chechik, Gal
        and Peng, Xue Bin
    },
    title = {
        MaskedMimic: Unified Physics-Based Character Control Through
        Masked Motion Inpainting
    },
    year = {2024},
    journal={ACM Transactions on Graphics (TOG)},
    publisher={ACM New York, NY, USA}
}

@article{mentzer2023finite,
  title={Finite scalar quantization: Vq-vae made simple},
  author={Mentzer, Fabian and Minnen, David and Agustsson, Eirikur and Tschannen, Michael},
  journal={arXiv preprint arXiv:2309.15505},
  year={2023}
}

@inproceedings{wan2024tlcontrol,
  title={Tlcontrol: Trajectory and language control for human motion synthesis},
  author={Wan, Weilin and Dou, Zhiyang and Komura, Taku and Wang, Wenping and Jayaraman, Dinesh and Liu, Lingjie},
  booktitle={European Conference on Computer Vision},
  pages={37--54},
  year={2024},
  organization={Springer}
}

@inproceedings{xie2024omnicontrol,
      title={OmniControl: Control Any Joint at Any Time for Human Motion Generation},
      author={Yiming Xie and Varun Jampani and Lei Zhong and Deqing Sun and Huaizu Jiang},
      booktitle={The Twelfth International Conference on Learning Representations},
      year={2024}
}

@inproceedings{chen2023executing,
  title={Executing your Commands via Motion Diffusion in Latent Space},
  author={Chen, Xin and Jiang, Biao and Liu, Wen and Huang, Zilong and Fu, Bin and Chen, Tao and Yu, Gang},
  booktitle={Proceedings of the IEEE/CVF Conference on Computer Vision and Pattern Recognition},
  pages={18000--18010},
  year={2023}
}

@inproceedings{fan2025gotozero,
      title={Go to Zero: Towards Zero-shot Motion Generation with Million-scale Data}, 
      author={Ke Fan and Shunlin Lu and Minyue Dai and Runyi Yu and Lixing Xiao and Zhiyang Dou and Junting Dong and Lizhuang Ma and Jingbo Wang},
      booktitle={Proceedings of the IEEE/CVF International Conference on Computer Vision (ICCV)},
      year={2025},
      eprint={2507.07095},
      archivePrefix={arXiv},
      primaryClass={cs.CV},
      url={https://arxiv.org/abs/2507.07095}, 
    }

@inproceedings{lu2025scamo,
  title={Scamo: Exploring the scaling law in autoregressive motion generation model},
  author={Lu, Shunlin and Wang, Jingbo and Lu, Zeyu and Chen, Ling-Hao and Dai, Wenxun and Dong, Junting and Dou, Zhiyang and Dai, Bo and Zhang, Ruimao},
  booktitle={Proceedings of the Computer Vision and Pattern Recognition Conference},
  pages={27872--27882},
  year={2025}
}

@inproceedings{meng2025rethinking,
  title={Rethinking Diffusion for Text-Driven Human Motion Generation: Redundant Representations, Evaluation, and Masked Autoregression},
  author={Meng, Zichong and Xie, Yiming and Peng, Xiaogang and Han, Zeyu and Jiang, Huaizu},
  booktitle={Proceedings of the Computer Vision and Pattern Recognition Conference},
  pages={27859--27871},
  year={2025}
}

@inproceedings{zhou2024emdm,
  title={Emdm: Efficient motion diffusion model for fast and high-quality motion generation},
  author={Zhou, Wenyang and Dou, Zhiyang and Cao, Zeyu and Liao, Zhouyingcheng and Wang, Jingbo and Wang, Wenjia and Liu, Yuan and Komura, Taku and Wang, Wenping and Liu, Lingjie},
  booktitle={European Conference on Computer Vision},
  pages={18--38},
  year={2024},
  organization={Springer}
}

@article{wu2025uniphys,
  title={UniPhys: Unified Planner and Controller with Diffusion for Flexible Physics-Based Character Control},
  author={Wu, Yan and Karunratanakul, Korrawe and Luo, Zhengyi and Tang, Siyu},
  journal={arXiv preprint arXiv:2504.12540},
  year={2025}
}

@article{huang2025diffusecloc,
  title={Diffuse-cloc: Guided diffusion for physics-based character look-ahead control},
  author={Huang, Xiaoyu and Truong, Takara and Zhang, Yunbo and Yu, Fangzhou and Sleiman, Jean Pierre and Hodgins, Jessica and Sreenath, Koushil and Farshidian, Farbod},
  journal={ACM Transactions on Graphics (TOG)},
  volume={44},
  number={4},
  pages={1--12},
  year={2025},
  publisher={ACM New York, NY, USA}
}

@inproceedings{karunratanakul2024optimizing,
  title={Optimizing diffusion noise can serve as universal motion priors},
  author={Karunratanakul, Korrawe and Preechakul, Konpat and Aksan, Emre and Beeler, Thabo and Suwajanakorn, Supasorn and Tang, Siyu},
  booktitle={Proceedings of the IEEE/CVF Conference on Computer Vision and Pattern Recognition},
  pages={1334--1345},
  year={2024}
}

@article{barquero2024flowmdm,
  title={Seamless Human Motion Composition with Blended Positional Encodings},
  author={Barquero, German and Escalera, Sergio and Palmero, Cristina},
  booktitle={Proceedings of the IEEE/CVF Conference on Computer Vision and Pattern Recognition},
  year={2024}
}

@inproceedings{petrovich24stmc,
    title     = {Multi-Track Timeline Control for Text-Driven 3D Human Motion Generation},
    author    = {Petrovich, Mathis and Litany, Or and Iqbal, Umar and Black, Michael J. and Varol, G{\"u}l and Peng, Xue Bin and Rempe, Davis},
    booktitle = {CVPR Workshop on Human Motion Generation},
    year      = {2024}
}

@inproceedings{pinyoanuntapong2024mmm,
  title={Mmm: Generative masked motion model},
  author={Pinyoanuntapong, Ekkasit and Wang, Pu and Lee, Minwoo and Chen, Chen},
  booktitle={Proceedings of the IEEE/CVF Conference on Computer Vision and Pattern Recognition},
  pages={1546--1555},
  year={2024}
}

@article{ling2020character,
  title={Character controllers using motion vaes},
  author={Ling, Hung Yu and Zinno, Fabio and Cheng, George and Van De Panne, Michiel},
  journal={ACM Transactions on Graphics (TOG)},
  volume={39},
  number={4},
  pages={40--1},
  year={2020},
  publisher={ACM New York, NY, USA}
}

@article{starke2022deepphase,
  title={Deepphase: Periodic autoencoders for learning motion phase manifolds},
  author={Starke, Sebastian and Mason, Ian and Komura, Taku},
  journal={ACM Transactions on Graphics (ToG)},
  volume={41},
  number={4},
  pages={1--13},
  year={2022},
  publisher={ACM New York, NY, USA}
}

@article{starke2019neural,
  title={Neural state machine for character-scene interactions},
  author={Starke, Sebastian and Zhang, He and Komura, Taku and Saito, Jun},
  journal={ACM Transactions on Graphics},
  volume={38},
  number={6},
  pages={178},
  year={2019},
  publisher={ACM}
}

@inproceedings{hassan2021stochastic,
  title={Stochastic scene-aware motion prediction},
  author={Hassan, Mohamed and Ceylan, Duygu and Villegas, Ruben and Saito, Jun and Yang, Jimei and Zhou, Yi and Black, Michael J},
  booktitle={Proceedings of the IEEE/CVF International Conference on Computer Vision},
  pages={11374--11384},
  year={2021}
}

@article{luo2023universal,
  title={Universal humanoid motion representations for physics-based control},
  author={Luo, Zhengyi and Cao, Jinkun and Merel, Josh and Winkler, Alexander and Huang, Jing and Kitani, Kris and Xu, Weipeng},
  journal={arXiv preprint arXiv:2310.04582},
  year={2023}
}

@article{holden2017phase,
  title={Phase-functioned neural networks for character control},
  author={Holden, Daniel and Komura, Taku and Saito, Jun},
  journal={ACM Transactions on Graphics (TOG)},
  volume={36},
  number={4},
  pages={1--13},
  year={2017},
  publisher={ACM New York, NY, USA}
}

@inproceedings{jiang2024autonomous,
  title={Autonomous character-scene interaction synthesis from text instruction},
  author={Jiang, Nan and He, Zimo and Wang, Zi and Li, Hongjie and Chen, Yixin and Huang, Siyuan and Zhu, Yixin},
  booktitle={SIGGRAPH Asia 2024 Conference Papers},
  pages={1--11},
  year={2024}
}

@inproceedings{rempe2023trace,
  title={Trace and pace: Controllable pedestrian animation via guided trajectory diffusion},
  author={Rempe, Davis and Luo, Zhengyi and Bin Peng, Xue and Yuan, Ye and Kitani, Kris and Kreis, Karsten and Fidler, Sanja and Litany, Or},
  booktitle={Proceedings of the IEEE/CVF Conference on Computer Vision and Pattern Recognition},
  pages={13756--13766},
  year={2023}
}

@inproceedings{rempe2021humor,
  title={Humor: 3d human motion model for robust pose estimation},
  author={Rempe, Davis and Birdal, Tolga and Hertzmann, Aaron and Yang, Jimei and Sridhar, Srinath and Guibas, Leonidas J},
  booktitle={Proceedings of the IEEE/CVF international conference on computer vision},
  pages={11488--11499},
  year={2021}
}

@inproceedings{zhang2024tedi,
  title={Tedi: Temporally-entangled diffusion for long-term motion synthesis},
  author={Zhang, Zihan and Liu, Richard and Hanocka, Rana and Aberman, Kfir},
  booktitle={ACM SIGGRAPH 2024 Conference Papers},
  pages={1--11},
  year={2024}
}

@inproceedings{petrovich22temos,
     title = {{TEMOS}: Generating diverse human motions from textual descriptions},
     author = {Petrovich, Mathis and Black, Michael J. and Varol, G{\"u}l},
     booktitle = {European Conference on Computer Vision ({ECCV})},
     year = {2022}
 }

@article{zhang2024motiondiffuse,
  title={Motiondiffuse: Text-driven human motion generation with diffusion model},
  author={Zhang, Mingyuan and Cai, Zhongang and Pan, Liang and Hong, Fangzhou and Guo, Xinying and Yang, Lei and Liu, Ziwei},
  journal={IEEE transactions on pattern analysis and machine intelligence},
  volume={46},
  number={6},
  pages={4115--4128},
  year={2024},
  publisher={IEEE}
}

@article{xiao2025motionstreamer,
  title={MotionStreamer: Streaming Motion Generation via Diffusion-based Autoregressive Model in Causal Latent Space},
  author={Xiao, Lixing and Lu, Shunlin and Pi, Huaijin and Fan, Ke and Pan, Liang and Zhou, Yueer and Feng, Ziyong and Zhou, Xiaowei and Peng, Sida and Wang, Jingbo},
  journal={arXiv preprint arXiv:2503.15451},
  year={2025}
}

@inproceedings{ji2025towards,
  title={Towards immersive human-x interaction: A real-time framework for physically plausible motion synthesis},
  author={Ji, Kaiyang and Shi, Ye and Jin, Zichen and Chen, Kangyi and Xu, Lan and Ma, Yuexin and Yu, Jingyi and Wang, Jingya},
  booktitle={Proceedings of the IEEE/CVF International Conference on Computer Vision},
  pages={10173--10183},
  year={2025}
}

@inproceedings{zhang2023generating,
  title={T2M-GPT: Generating Human Motion from Textual Descriptions with Discrete Representations},
  author={Zhang, Jianrong and Zhang, Yangsong and Cun, Xiaodong and Huang, Shaoli and Zhang, Yong and Zhao, Hongwei and Lu, Hongtao and Shen, Xi},
  booktitle={Proceedings of the IEEE/CVF Conference on Computer Vision and Pattern Recognition (CVPR)},
  year={2023},
}

@article{jiang2024motiongpt,
  title={Motiongpt: Human motion as a foreign language},
  author={Jiang, Biao and Chen, Xin and Liu, Wen and Yu, Jingyi and Yu, Gang and Chen, Tao},
  journal={Advances in Neural Information Processing Systems},
  volume={36},
  year={2024}
}

@inproceedings{cen2025ready_to_react,
  title={Ready-to-React: Online Reaction Policy for Two-Character Interaction Generation},
  author={Cen, Zhi and Pi, Huaijin and Peng, Sida and Shuai, Qing and Shen, Yujun and Bao, Hujun and Zhou, Xiaowei and Hu, Ruizhen},
  booktitle={ICLR},
  year={2025}
}

@inproceedings{li2025unimotion,
  title={Unimotion: Unifying 3d human motion synthesis and understanding},
  author={Li, Chuqiao and Chibane, Julian and He, Yannan and Pearl, Naama and Geiger, Andreas and Pons-Moll, Gerard},
  booktitle={2025 International Conference on 3D Vision (3DV)},
  pages={240--249},
  year={2025},
  organization={IEEE}
}

@inproceedings{karunratanakul2023gmd,
  title={Guided motion diffusion for controllable human motion synthesis},
  author={Karunratanakul, Korrawe and Preechakul, Konpat and Suwajanakorn, Supasorn and Tang, Siyu},
  booktitle={Proceedings of the IEEE/CVF International Conference on Computer Vision},
  pages={2151--2162},
  year={2023}
}

@article{liao2025beyondmimic,
  title={Beyondmimic: From motion tracking to versatile humanoid control via guided diffusion},
  author={Liao, Qiayuan and Truong, Takara E and Huang, Xiaoyu and Tevet, Guy and Sreenath, Koushil and Liu, C Karen},
  journal={arXiv preprint arXiv:2508.08241},
  year={2025}
}

@article{he2025asap,
  title={Asap: Aligning simulation and real-world physics for learning agile humanoid whole-body skills},
  author={He, Tairan and Gao, Jiawei and Xiao, Wenli and Zhang, Yuanhang and Wang, Zi and Wang, Jiashun and Luo, Zhengyi and He, Guanqi and Sobanbab, Nikhil and Pan, Chaoyi and others},
  journal={arXiv preprint arXiv:2502.01143},
  year={2025}
}

@article{zhao2025resmimic,
  title={ResMimic: From General Motion Tracking to Humanoid Whole-body Loco-Manipulation via Residual Learning},
  author={Zhao, Siheng and Ze, Yanjie and Wang, Yue and Liu, C Karen and Abbeel, Pieter and Shi, Guanya and Duan, Rocky},
  journal={arXiv preprint arXiv:2510.05070},
  year={2025}
}

@article{
	cohan2024condmdi,
	author = {Setareh, Cohan and Tevet, Guy and Reda, Daniele and Peng, Xue Bin and van de Panne, Michiel},
	title = {Flexible Motion In-betweening with Diffusion Models},
	year = {2024},
	publisher = {Association for Computing Machinery},
	address = {New York, NY, USA},
	booktitle = {ACM SIGGRAPH 2024 Conference Proceedings},
	location = {Los Angeles, CA, USA},
	series = {SIGGRAPH '24}
}

@InProceedings{guo2022humanml3d,
    author    = {Guo, Chuan and Zou, Shihao and Zuo, Xinxin and Wang, Sen and Ji, Wei and Li, Xingyu and Cheng, Li},
    title     = {Generating Diverse and Natural 3D Human Motions From Text},
    booktitle = {Proceedings of the IEEE/CVF Conference on Computer Vision and Pattern Recognition (CVPR)},
    month     = {June},
    year      = {2022},
    pages     = {5152-5161}
}

@inproceedings{zhang2025primal,
  author = {Zhang, Yan and Feng, Yao and Cseke, Alpár and Saini, Nitin and Bajandas, Nathan and Heron, Nicolas and Black, Michael J.},  
   title = {{PRIMAL:} Physically Reactive and Interactive Motor Model for Avatar Learning},
  booktitle = {Proceedings of the IEEE/CVF International Conference on Computer Vision (ICCV)},
  month = oct,
  year = {2025}
}

@article{ren2023insactor,
  title={InsActor: Instruction-driven Physics-based Characters},
  author={Ren, Jiawei and Zhang, Mingyuan and Yu, Cunjun and Ma, Xiao and Pan, Liang and Liu, Ziwei},
  journal={NeurIPS},
  year={2023}
}

@inproceedings{fragkiadaki2015recurrent,
  title={Recurrent network models for human dynamics},
  author={Fragkiadaki, Katerina and Levine, Sergey and Felsen, Panna and Malik, Jitendra},
  booktitle={Proceedings of the IEEE international conference on computer vision},
  pages={4346--4354},
  year={2015}
}

@article{taylor2006modeling,
  title={Modeling human motion using binary latent variables},
  author={Taylor, Graham W and Hinton, Geoffrey E and Roweis, Sam},
  journal={Advances in neural information processing systems},
  volume={19},
  year={2006}
}

@inproceedings{pinyoanuntapong2024bamm,
  title={Bamm: Bidirectional autoregressive motion model},
  author={Pinyoanuntapong, Ekkasit and Saleem, Muhammad Usama and Wang, Pu and Lee, Minwoo and Das, Srijan and Chen, Chen},
  booktitle={European Conference on Computer Vision},
  pages={172--190},
  year={2024},
  organization={Springer}
}

@article{plappert2016kit,
  title={The kit motion-language dataset},
  author={Plappert, Matthias and Mandery, Christian and Asfour, Tamim},
  journal={Big data},
  volume={4},
  number={4},
  pages={236--252},
  year={2016},
  publisher={Mary Ann Liebert, Inc. 140 Huguenot Street, 3rd Floor New Rochelle, NY 10801 USA}
}

@article{radford2018gpt,
  title={Improving language understanding by generative pre-training},
  author={Radford, Alec and Narasimhan, Karthik and Salimans, Tim and Sutskever, Ilya and others},
  year={2018},
  publisher={San Francisco, CA, USA}
}

@inproceedings{zhao2023synthesizing,
  title={Synthesizing diverse human motions in 3d indoor scenes},
  author={Zhao, Kaifeng and Zhang, Yan and Wang, Shaofei and Beeler, Thabo and Tang, Siyu},
  booktitle={Proceedings of the IEEE/CVF international conference on computer vision},
  pages={14738--14749},
  year={2023}
}

@inproceedings{zhang2022wanderings,
  title={The wanderings of odysseus in 3d scenes},
  author={Zhang, Yan and Tang, Siyu},
  booktitle={Proceedings of the IEEE/CVF Conference on Computer Vision and Pattern Recognition},
  pages={20481--20491},
  year={2022}
}

@article{peng2022ase,
  title={Ase: Large-scale reusable adversarial skill embeddings for physically simulated characters},
  author={Peng, Xue Bin and Guo, Yunrong and Halper, Lina and Levine, Sergey and Fidler, Sanja},
  journal={ACM Transactions On Graphics (TOG)},
  volume={41},
  number={4},
  pages={1--17},
  year={2022},
  publisher={ACM New York, NY, USA}
}

@inproceedings{zhou2019continuity,
  title={On the continuity of rotation representations in neural networks},
  author={Zhou, Yi and Barnes, Connelly and Lu, Jingwan and Yang, Jimei and Li, Hao},
  booktitle={Proceedings of the IEEE/CVF conference on computer vision and pattern recognition},
  pages={5745--5753},
  year={2019}
}

@inproceedings{
    reddy2024llm2vec,
    title={{LLM2V}ec: Large Language Models Are Secretly Powerful Text Encoders},
    author={Parishad BehnamGhader and Vaibhav Adlakha and Marius Mosbach and Dzmitry Bahdanau and Nicolas Chapados and Siva Reddy},
    booktitle={First Conference on Language Modeling},
    year={2024},
    url={https://openreview.net/forum?id=IW1PR7vEBf}
}

@inproceedings{guo2020action2motion,
  title={Action2motion: Conditioned generation of 3d human motions},
  author={Guo, Chuan and Zuo, Xinxin and Wang, Sen and Zou, Shihao and Sun, Qingyao and Deng, Annan and Gong, Minglun and Cheng, Li},
  booktitle={Proceedings of the 28th ACM international conference on multimedia},
  pages={2021--2029},
  year={2020}
}

@inproceedings{kingma2014auto,
  title={Auto-Encoding Variational Bayes},
  author={Kingma, Diederik P and Welling, Max},
  booktitle={International Conference on Learning Representations},
  year={2014}
}

@INPROCEEDINGS{girshickFastRCNN2015,

  author={Girshick, Ross},

  booktitle={International Conference on Computer Vision (ICCV)}, 

  title={{Fast R-CNN}}, 

  year={2015},

}

@inproceedings{hoClassifierFreeDiffusionGuidance2022,
  title={Classifier-Free Diffusion Guidance},
  author={Ho, Jonathan and Salimans, Tim},
  booktitle={NeurIPS 2021 Workshop on Deep Generative Models and Downstream Applications},
  year={2021}
}

@inproceedings{petrovich23tmr,
    title     = {{TMR}: Text-to-Motion Retrieval Using Contrastive {3D} Human Motion Synthesis},
    author    = {Petrovich, Mathis and Black, Michael J. and Varol, G{\"u}l},
    booktitle = {International Conference on Computer Vision ({ICCV})},
    year      = {2023}
}

@article{yi2025viser,
  title={Viser: Imperative, web-based 3d visualization in python},
  author={Yi, Brent and Kim, Chung Min and Kerr, Justin and Wu, Gina and Feng, Rebecca and Zhang, Anthony and Kulhanek, Jonas and Choi, Hongsuk and Ma, Yi and Tancik, Matthew and Kanazawa, Angjoo},
  journal={arXiv preprint arXiv:2507.22885},
  year={2025}
}

@inproceedings{
geng2025mean,
title={Mean Flows for One-step Generative Modeling},
author={Zhengyang Geng and Mingyang Deng and Xingjian Bai and J Zico Kolter and Kaiming He},
booktitle={The Thirty-ninth Annual Conference on Neural Information Processing Systems},
year={2025},
}

@inproceedings{
lu2025simplifying,
title={Simplifying, Stabilizing and Scaling Continuous-time Consistency Models},
author={Cheng Lu and Yang Song},
booktitle={The Thirteenth International Conference on Learning Representations},
year={2025},
}

@article{SMPL:2015,
title = {{SMPL}: A Skinned Multi-Person Linear Model},
author = {Loper, Matthew and Mahmood, Naureen and Romero, Javier and Pons-Moll, Gerard and Black, Michael J.},
journal = {ACM Trans. Graphics (Proc. SIGGRAPH Asia)},
year = {2015}
}

@article{ho2020denoising,
  title={Denoising diffusion probabilistic models},
  author={Ho, Jonathan and Jain, Ajay and Abbeel, Pieter},
  journal={Advances in neural information processing systems},
  volume={33},
  pages={6840--6851},
  year={2020}
}

@article{llama3modelcard,
title={Llama 3 Model Card},
author={AI@Meta},
year={2024},
url = {https://github.com/meta-llama/llama3/blob/main/MODEL_CARD.md}
}

@article{everett2024scaling,
  title={Scaling exponents across parameterizations and optimizers},
  author={Everett, Katie and Xiao, Lechao and Wortsman, Mitchell and Alemi, Alexander A and Novak, Roman and Liu, Peter J and Gur, Izzeddin and Sohl-Dickstein, Jascha and Kaelbling, Leslie Pack and Lee, Jaehoon and others},
  journal={International Conference on Machine Learning},
  year={2024}
}

@article{holden2020learned,
  title={Learned motion matching},
  author={Holden, Daniel and Kanoun, Oussama and Perepichka, Maksym and Popa, Tiberiu},
  journal={ACM Transactions on Graphics (ToG)},
  year={2020},
}

@article{gou2025control,
  title={Control Operators for Interactive Character Animation},
  author={Gou, Ruiyu and van de Panne, Michiel and Holden, Daniel},
  journal={ACM Transactions on Graphics (TOG)},
  year={2025},
}

@article{Kimodo2026,
  title={Kimodo: Scaling Controllable Human Motion Generation},
  author={Rempe, Davis and Petrovich, Mathis and Yuan, Ye and Zhang, Haotian and Peng, Xue Bin and Jiang, Yifeng and Wang, Tingwu and Iqbal, Umar and Minor, David and de Ruyter, Michael and Li, Jiefeng and Tessler, Chen and Lim, Edy and Jeong, Eugene and Wu, Sam and Hassani, Ehsan and Huang, Michael and Yu, Jin-Bey and Chung, Chaeyeon and Song, Lina and Dionne, Olivier and Kautz, Jan and Yuen, Simon and Fidler, Sanja},
  journal={arXiv:2603.15546},
  year={2026}
}

@article{luo2025sonic,
    title={SONIC: Supersizing Motion Tracking for Natural Humanoid Whole-Body Control},
    author={Luo, Zhengyi and Yuan, Ye and Wang, Tingwu and Li, Chenran and Chen, Sirui and Casta\~neda, Fernando and Cao, Zi-Ang and Li, Jiefeng and Minor, David and Ben, Qingwei and Da, Xingye and Ding, Runyu and Hogg, Cyrus and Song, Lina and Lim, Edy and Jeong, Eugene and He, Tairan and Xue, Haoru and Xiao, Wenli and Wang, Zi and Yuen, Simon and Kautz, Jan and Chang, Yan and Iqbal, Umar and Fan, Linxi and Zhu, Yuke},
    journal={arXiv preprint arXiv:2511.07820},
    year={2025}
}

@misc{BonesStudio2026,
  author        = {{Bones Studio}},
  title        = {{AI} Datasets for Machine Learning and Motion Capture},
  howpublished = {\url{https://bones.studio/datasets}},
  year         = {2026},
  note         = {Accessed: 2026}
}

\end{document}